# A century of coherent states

**Dušan POPOV [a, b]**

[a] University Politehnica Timisoara, Department of Physical Foundations of Engineering, B-dul Vasile Pârvan No. 2, 300223 Timisoara, Romania
[b] Serbian Academy of Nonlinear Sciences (SANS), Kneza Mihaila 36, Beograd-Stari Grad, Belgrade, Serbia
E-mail: dusan_popov@yahoo.co.uk
ORCID: https://orcid.org/0000-0003-3631-3247

**Abstract**

During the century of existence of the notion of coherent states, either linear or nonlinear, several schemes for their construction, theoretical or experimental, have been developed. Generally, the mathematical structure of coherent states depends on the choice of ladder operators, and consequently on the structure constants. In this paper, we propose a way to construct generalized coherent states for anharmonic oscillators that is based on a diagonal operator ordering technique (DOOT) applied to generalized hypergeometric functions, that is, on some of the most general special functions. These states are generated by the action of a pair of ladder operators, the creation $\hat{\mathcal{A}}_{+}$ and the annihilation $\hat{\mathcal{A}}_{-}$, whose ordered normal product is equal to the dimensionless Hamiltonian $\hat{\mathcal{H}}$ of the quantum system. In addition, the action of these operators is easy to find if the expression for the dimensionless energy eigenvalues $e(n)$ is known.

## 1. Introduction

This year, 2026, marks exactly a century since the concept of “coherent states” was introduced into quantum mechanics. This fact is due to Erwin Schrödinger who, in his work [Schrödinger, 1926], wanted to find solutions of the equation (which would later bear his name) that would satisfy the correspondence principle. In other words, to find, for the quantum harmonic oscillator, such states whose temporal behavior would be as close as possible to the behavior of the quantum harmonic oscillator. Although Schrödinger did not use the notion of "coherent states" in this paper, the scientific community nevertheless attributed to him the paternity of introducing this notion. Since the coherent state describes (or is equivalent to) a state of a particle whose associated fundamental wave packet moves with respect to the origin of the coordinate system, then the connection with the classical solutions becomes realized if it is considered that the particle oscillates with an amplitude equivalent to its displacement. Subsequently, for several decades, this notion seemed to be forgotten, it did not arouse greater interest among researchers. It seems that the development of laser radiation techniques and applications had a great influence on the reactivation of the notion of coherent states. The one who reactivated (practically, reintroduced) this notion, considering it as a true bridge between classical and quantum mechanics, was Roy Glauber, in the early 1960s, when he used coherent states to explain the coherence of the electromagnetic field in quantum optics, respectively related to the problem of laser beam coherence [Glauber, 1963]. Implicitly, these states being related to the phenomenon of coherence of laser radiation, Glauber called them "coherent states" and defined them mathematically as eigenvectors of the annihilation operator of the harmonic oscillator. [1]

In the same period, the contributions of J. R. Klauder [Klauder, 1963] are no less important for the applications of the coherent states formalism. Not long after, Barut and Girardello constructed generalized coherent states defined as the eigenstates of the lowering and raising operators of non-compact groups [Barut, 1971]. Around the same time, to bridge the gap between classical and quantum mechanics, coherent states for systems possessing specific symmetry groups were constructed based on the action of the generators of the SU(1, 1) and SU(2) quantum groups associated with the studied systems [Perelomov, 1972], [Perelomov, 1986]. A change of attitude regarding the construction of coherent states occurred with the consideration of the properties of generalized hypergeometric functions, ${}_pF_q\left(\ldots;\ldots;|z|^2\right)$, with $p$ and $q$ - non negative integers [Appl, 2004], [Popov, 2015b]. Depending on the region in the complex plane $z$ on which they are defined, generalized hypergeometric functions can be classified into three categories: a) those defined on the entire complex space, $|z|<\infty$; b) those defined inside a circle with the center at the origin of the coordinate system and radius equal to convergence radius $R_c$, $|z|<R_c$; and c) those defined on a circle, usually of convergence radius equal to unity $|z|=1$.This tacitly waived, in the case of nonlinear coherent states, the initial condition that coherent states minimize uncertainty relations. This implicitly broadened the area of construction and applicability of coherent states. Later, in recent decades, the degree of generalization of coherent states diversified. Thus, coherent states connected with different generalized functions (Mittag-Leffler, Fox-Wright, etc.) appeared which brought, through their possible applications [Antoine, 2001], [Klauder, 2001], [Dattoli, 2017], [Popov, 2024], [Giraldi, 2025], [Droghei, 2025]. During the decades, the rules of definition were correspondingly changed [Klauder, 2001a] because the mathematical structure of coherent states depends crucial on the choice of creation and annihilation operators, that is, in concrete terms, on the result of their action on the Fock vectors. The choise of these operators directly changes the mathematical form of the coherent states, their overcompleteness, and their stability. Therefore, in our approach we have chosen such pair of operators, $\hat{\mathcal{A}}_+$ and the annihilation $\hat{\mathcal{A}}_-$, whose ordered normal product is even equal to the dimensionless Hamiltonian $\hat{\mathcal{H}}$ of the quantum system.

The present paper has the following structure: In Sec. 2 we briefly and comparatively refer to linear (one-dimensional harmonic oscillator) and nonlinear quantum systems, highlighting the transition / connection between their coherent states. In Sec. 3 we first briefly presented the rules of the creation and annihilation operator ordering techniques, IWOP (Integration Within an Ordered Product) and DOOT (Diagonal Operators Ordered Technique). By specifically choosing the generalized expression of the energy eigenvalues, ${}_pe_q(n)$, these rules allow the construction of generalized coherent states for nonlinear systems. We applied the

[1] For his contributions related to the quantum theory of optical coherence, Roy Glauber received the Nobel Prize in Physics in 2005. https://www.nobelprize.org/prizes/physics/2005/glauber/lecture/

DOOT technique to construct generalized coherent states starting from the three "traditional" definitions (Barut-Girardello, Klauder-Perelomov and Gazeau-Klauder). In Sec. 4 we show how the DOOT technique can be used in the case of mixed (thermal) states. Following the same ideas, Sec. 5 was dedicated to the construction of coherent states for continuous spectrum systems, also using the DOOT technique, and in Sec. 6 we applied the previous considerations, for the case of several quantum oscillators, by which the corresponding generalized coherent states are directly obtained. In Sec. 7 we present some concluding remarks regarding the generalized coherent states.

## 2. Linear and nonlinear systems

### 2.1 Briefly about the linear harmonic oscillator

From quantum mechanics books it is known that the linear harmonic oscillator's Hamiltonian can also be expressed through the creation $\hat{\boldsymbol{a}}^+$ and annihilation $\hat{\boldsymbol{a}}$ operators which fulfill the canonical commutation relation $\left[\hat{\boldsymbol{a}},\, \hat{\boldsymbol{a}}^+\right]=1$:

$$\hat{\mathcal{H}}_{HO-1D}(\hat{\boldsymbol{a}}^+,\, \hat{\boldsymbol{a}}) = \hbar\omega\left(\hat{\boldsymbol{a}}^+\hat{\boldsymbol{a}} + \frac{1}{2}\right) \tag{2.1}$$

the corresponding Schrödinger equation for stationary states being $\hat{\mathcal{H}}_{HO-1D}\,|\,n> = E_n\,|\,n>$

The Fock - vectors $|\,n>$ in the special infinite dimensional Hilbert space (the Fock space) are orthogonal $<n\,|\,m> = \delta_{nm}$ forming a vector basis $\left\{|\,n>, n = 0,1,2,\ldots\right\}$.

The boson operators act on the Fock vectors in the following manner:

$$\begin{aligned} &\hat{\boldsymbol{a}}^+\,|\,n> = \sqrt{n+1}\,|\,n+1>, \quad \hat{\boldsymbol{a}}\,|\,n> = \sqrt{n}\,|\,n-1>, \quad \hat{\boldsymbol{a}}^+\hat{\boldsymbol{a}}\,|\,n> = n\,|\,n> \\ &<n\,|\,\hat{\boldsymbol{a}}^+ = \sqrt{n}<n-1\,|, \quad <n\,|\,\hat{\boldsymbol{a}} = \sqrt{n+1}<n+1\,|, \; <n\,|\,\hat{\boldsymbol{a}}^+\hat{\boldsymbol{a}} = n<n\,| \end{aligned} \tag{2.2}$$

Two important observations are required: 1.) Fock vectors $|n>$ are not eigenvectors of the creation $\hat{\boldsymbol{a}}^+$ and annihilation $\hat{\boldsymbol{a}}$ operators; 2.) Fock vectors are eigenvectors of the product $\hat{\boldsymbol{a}}^+\hat{\boldsymbol{a}}$, i.e. of the particle number operator $\hat{\mathcal{N}} \equiv \hat{\boldsymbol{a}}^+\hat{\boldsymbol{a}}$, where $\hat{\mathcal{N}}|n> \equiv \hat{\boldsymbol{a}}^+\hat{\boldsymbol{a}}|n> = n|n>$. Consequently, the vectors $|\,n>$ are the eigenvectors of the Hamiltonian:

$$\hat{\mathcal{H}}_{HO-1D}\,|\,n> = \hbar\omega\left(n + \frac{1}{2}\right)|\,n> \tag{2.3}$$

where $n$ is the main quantum number.

### 2.2 Nonlinear quantum systems

In most cases, natural processes have irreversible and non-linear character. Consequently, from a physical point of view, linearity is only a first degree of approximation in infinitely examination of nature, because causes and the consequences are not always proportional in magnitude. In quantum mechanics, the superposition principle is valid, according to which, if two solutions satisfy the conditions of the problem, then their linear combination is also a correct solution. However, experimental measurements show that in practice this does not happen always, so such a combination often leads to irregular or physically unacceptable solutions. On the other hand, when examining nonlinear phenomena, the linear approach has a justification only as a first-order approximation, which is not sufficient for a more or less complete description of the system's behavior. The Schrödinger equation being a linear equation, provides a sufficient description for some phenomena. Nonlinear aspects begin when various nonlinear variants of this equation are used. However, nonlinearities in quantum mechanics also appear in other situations, in which the Schrödinger equation does not intervene directly.

As a first example, let's refer to oscillatory phenomena. The cause of oscillations, a conservative force $F(x)$, i.e. the negative gradient of the potential energy $U(x)$, can be developed in power series around the equilibrium position $x_0 = 0$:

$$F(x) = -\frac{dU(x)}{dx} \quad \underbrace{\underbrace{= \underbrace{F(0)}_{\substack{\text{constant} \\ =0}} + \underbrace{\frac{1}{1!}\left(\frac{dF}{dx}\right)_{x=0}}_{=0} x \quad + \frac{1}{2!}\underbrace{\left(\frac{d^2F}{dx^2}\right)_{x=0}}_{=-k_e}}_{\text{elastical force - cause of harmonic oscillations}} x^2 + \frac{1}{3!}\left(\frac{d^3F}{dx^3}\right)_{x=0} x^3 + \ldots}_{\text{non elasic force - cause of anharmonic oscillations}} \quad , \tag{2.4}$$

where you can see that its *linear component* of the elastic type is responsible for *harmonic oscillations*. The corresponding potential energy can also be developed in power series around the equilibrium point:

$$\underbrace{\underbrace{U(x) = \underbrace{U(0)}_{\substack{\text{constant} \\ =0}} + \underbrace{x\frac{dU}{dx}\bigg|_{x=0}}_{=0} + \frac{x^2}{2!}\underbrace{\frac{d^2U}{dx^2}\bigg|_{x=0}}_{k_e}}_{\text{potential energy of the harmonic oscillations = purely paraboloc}} + \frac{x^3}{3!}\frac{d^3U}{dx^3}\bigg|_{x=0}}_{\text{potential energy of the anharmonic oscillations = non purely parabolic}} + \ldots \equiv \frac{k_e}{2}x^2 + O(x^3) \tag{2.5}$$

where the square term corresponds to the potential energy of *harmonic oscillations*. The function $O(x^3)$ can be considered as the *anharmonic perturbation function* or the *non linearity function*.

So, we can observe that between the linear / nonlinear behavior of a certain physical system and its harmonic / anharmonic oscillations there are the following ***similarities***:

***linear behavior*** $\Leftrightarrow$ ***harmonic oscillations***

***nonlinear behavior*** $\Leftrightarrow$ ***anharmonic oscillations***

Moreover, the behavior of the one-dimensional harmonic oscillator is described by a linear differential equation, in which the restoring force is directly proportional to the displacement from the equilibrium position. Therefore, the harmonic oscillator is considered a linear system. In contrast, other quantum oscillator systems (Morse, Pöschl-Teller, etc.) are anharmonic oscillators, their potentials not being purely parabolic. These oscillators are considered to be nonlinear systems. More briefly, the expression of potential energy can constitute the first criterion for classifying systems, briefly called the ***criterion of potential energy*** and expressed schematically as follows:

$$U(x)=\begin{cases}\frac{k_e}{2}x^2 & \rightarrow \quad \text{linear system} \quad \rightarrow \quad \text{linear coherent states}\\ \frac{k_e}{2}x^2+O(x^3) & \rightarrow \quad \text{nonlinear systems} \quad \rightarrow \quad \text{nonlinear coherent states}\end{cases} \tag{2.6}$$

Another criterion for establishing the linearity / non-linearity of a system is based on the expression of the eigenvalues of the system's energy, specifically its dependence on the principal or main quantum number $n$. We will call it the ***energy eigenvalues criterion***.

The eigenvalues of the energy of the one-dimensional harmonic oscillator depend linearly on $n$, that is $E_n=\hbar\omega\left(n+\frac{1}{2}\right)$. This implies that the Hamiltonian $\hat{\mathcal{H}}$ of the system also depends linearly on the particle number operator $\hat{\mathcal{N}}$, that is, on the product $\hat{a}^+\hat{a}$.

Generally the Hamiltonian of a certain quantum system must have such algebraic structure that contains only the entire degrees of particle number operator $\hat{\mathcal{N}}$, i.e. the products of kind $(\hat{a}^+\hat{a})^s$, $s=1,2,...$, and not otherwise combinations of operators $\hat{a}^+$ and $\hat{a}$, because, for these combinations, the vectors $|n>$ are not the eigenvectors of the Hamiltonian. The set of operators $\{\hat{\mathcal{N}},\ \hat{a}^+,\ \hat{a},\ \hat{\mathcal{I}}\}$ spans a Lie algebra $h_4$ and the corresponding Lie group is the Heisenberg-Weyl group $H_4$ [Zhang, 1990].

For a nonlinear system the Hamiltonian will be a nonlinear function of $\hat{\mathcal{N}}$, respectively $\hat{a}^+\hat{a}$ and we can expand it in power series (which can be interrupted at a certain power, becoming a polynomial in $n$).

$$\hat{\mathcal{H}}=\hat{\mathcal{H}}\left(\hat{\mathcal{N}}\right)=\sum_{j=0}^{j_{\max}\le\infty}c_j\hat{\mathcal{N}}^j \quad \Leftrightarrow \quad \hat{\mathcal{H}}=\hat{\mathcal{H}}\left(\hat{a}^+\hat{a}\right)=\sum_{j=0}^{j_{\max}\le\infty}c_j\,\#\left(\hat{a}^+\hat{a}\right)^j\# \tag{2.7}$$

$$\hat{\mathcal{H}}\left(\hat{\mathcal{N}}\right)|\,n>=\sum_{j=0}^{j_{\max}\le\infty}c_j\hat{\mathcal{N}}^j\,|\,n>=\sum_{j=0}^{j_{\max}\le\infty}c_jn^j\,|\,n>\ ,\ E_n=c_0+c_1\,n+\sum_{j=2}^{j_{\max}\le\infty}c_jn^j \tag{2.8}$$

Nonlinear effects appear starting with the term of the second power of $n$. The expression of the energy eigenvalues will be written, therefore

$$E_n=c_0+c_1\,n+\sum_{j=2}^{j_{\max}\le\infty}c_jn^j=c_0+n\left(c_1+c_1\,n+\sum_{j=2}^{j_{\max}\le\infty}c_jn^{j-1}\right) \tag{2.9}$$

Using the notations $c_1+c_1\,n+\sum_{j=2}^{j_{\max}\le\infty}c_jn^{j-1}\equiv\left[f(n)\right]^2$, $E_n\equiv\hbar\,\omega\;e(n), c_0\equiv\hbar\,\omega\;e(0)$, the dimensionless energy eigenvalues becomes

$$e(n)=e(0)+n\left[f(n)\right]^2 \tag{2.10}$$

Therefore, the classification criterion of the second system, the ***energy eigenvalue criterion***, will be written, in condensed form, as follows:

$$e(n)-e(0)=\begin{cases} n & \rightarrow \quad \text{linear system}\\ n\left[f(n)\right]^2 & \rightarrow \quad \text{non linear system}\end{cases} \tag{2.11}$$

From here it follows that it is logical to introduce a new pair of non Hermitian ladder operators, the creation ${}_p\hat{\mathcal{A}}_{q,+}$ and the ${}_p\hat{\mathcal{A}}_{q,-}$ annihilation operators. For brevity, we will skip writing the indices $p$ and $q$, so we will simply write ${}_p\hat{\mathcal{A}}_{q,+} = \hat{\mathcal{A}}_+$ and ${}_p\hat{\mathcal{A}}_{q,-} = \hat{\mathcal{A}}_-$ . These are adjoint operators, $\left(\hat{\mathcal{A}}_+\right)^+ = \hat{\mathcal{A}}_-$ and $\left(\hat{\mathcal{A}}_-\right)^+ = \hat{\mathcal{A}}_+$, but not self-adjoint, that is, they are not Hermitian. Also, these operators are not linear, and are connected with the canonical operators $\hat{a}^+$ and $\hat{a}$ through a function ${}_p f_q\left(\hat{\mathcal{N}}\right)$ depending of the particle number operator $\hat{\mathcal{N}} = \hat{a}^+\hat{a}$ . This real function ${}_p f_q\left(\hat{\mathcal{N}}\right)$ is called the nonlinearity function, where $p$ and $q$ are positive integer numbers. Consequently, the new introduced non linear operators have the following structure

$$\hat{\mathcal{A}}_- = \hat{a}\ {}_p f_q\left(\hat{\mathcal{N}}\right) \quad , \quad \hat{\mathcal{A}}_+ = {}_p f_q\left(\hat{\mathcal{N}}\right)\hat{a}^+ \tag{2.12}$$

The pair of creation / annihilation operators above was not chosen randomly, but so that their normal ordered product is exactly equal to the Hamiltonian of the system, $\hat{\mathcal{H}} = \hat{\mathcal{A}}_+\hat{\mathcal{A}}_-$ .

Their normal ordered product is

$$\#\hat{\mathcal{A}}_+\hat{\mathcal{A}}_-\# = \#{}_p f_q\left(\hat{\mathcal{N}}\right)\hat{a}^+\hat{a}\ {}_p f_q\left(\hat{\mathcal{N}}\right)\# = \#\hat{a}^+\hat{a}\left[{}_p f_q\left(\hat{\mathcal{N}}\right)\right]^2\# \tag{2.13}$$

and the actions of these operators and their dual on the Fock vectors are

$$\hat{\mathcal{A}}_- \,|\, n> = \sqrt{{}_p e_q(n)}\,|\, n-1> \quad , \quad <n\,|\,\hat{\mathcal{A}}_+ = \sqrt{{}_p e_q(n)} < n-1\,| \ ,$$
$$\hat{\mathcal{A}}_+ \,|\, n> = \sqrt{{}_p e_q(n+1)}\,|\, n+1> , \quad <n\,|\,\hat{\mathcal{A}}_- = \sqrt{{}_p e_q(n+1)} < n+1\,| \tag{2.14}$$

$$<n\,|\,\hat{\mathcal{A}}_+\hat{\mathcal{A}}_-\,|\,n> = {}_p e_q(n) \quad , \quad <n\,|\,\hat{\mathcal{A}}_-\hat{\mathcal{A}}_+\,|\,n> = {}_p e_q(n+1)$$
$$<n\,|\left[\hat{\mathcal{A}}_-\,,\,\hat{\mathcal{A}}_+\right]|\,n> = {}_p e_q(n+1) - {}_p e_q(n) , \quad <n\,|\left(\hat{\mathcal{A}}_-\hat{\mathcal{A}}_+ + \hat{\mathcal{A}}_+\hat{\mathcal{A}}_-\right)|\,n> = {}_p e_q(n+1) + {}_p e_q(n) \tag{2.15}$$

It can be seen that ${}_p e_q(n)$ are the dimensionless eigenvalues of the energy of the system.

The following relations are valid

$$|\,n> = \frac{1}{\sqrt{{}_p \rho_q(n)}}\left(\hat{\mathcal{A}}_+\right)^n |\,0> \quad , \quad <n\,| = \frac{1}{\sqrt{{}_p \rho_q(n)}} <0\,|\left(\hat{\mathcal{A}}_-\right)^n \tag{2.16}$$

where we used the notation

$${}_p \rho_q(n) = \prod_{j=0}^{n} {}_p e_q(j) \tag{2.17}$$

## 3. Generalized coherent states

### 3.1 IWOP and DOOT

In quantum mechanics, a major role is played by the creation / raising and the annihilation / lowering operators, commonly called ladder operators, as well as their product. There are three ways to order a product of two such operators: normal, antinormal and symmetric. Starting with the year 1983, Hong-Yi Fan and collaborators published a series of papers in which, for the first time, they used the calculations which were performed using the idea that the boson operators $\hat{a}$ and $\hat{a}^+$ commute with each other within ordered product, which is itself a nature of ordered products [Fan, 1983]. This idea later materialized into what was called *Integration Within an Ordered Product* (IWOP) [Fan, 1996], and with its help a series of remarkable results were obtained for calculations in quantum optics. IWOP is a technique for determining normally ordered forms of unitary operators, particularly those that induce symplectic transformations. Using this technique, calculations involving a normal ordering of operators are greatly simplified, especially within the CSs formalism.

Despite all the remarkable results obtained using the IWOP technique, it involved only the canonical creation $\hat{\boldsymbol{a}}^+$ and annihilation $\hat{\boldsymbol{a}}$ operators, characteristic for the linear harmonic oscillator. Therefore, the next logical step was to see if such a type of operational technique could not be applied to a corresponding pair of nonlinear operators, $\hat{\mathcal{A}}_+ = f\left(\hat{\mathcal{N}}\right)\hat{\boldsymbol{a}}^+$ and $\hat{\mathcal{A}}_- = \hat{\boldsymbol{a}}\, f\left(\hat{\mathcal{N}}\right)$. In this sense, we have maintained the rules formulated for IWOP, adapting them for problems involving linear operators. However, to distinguish the two techniques, we have named the new technique the *Diagonal Ordering Operation Technique* (DOOT), and instead of the pair of signs $\vdots \ldots \vdots$ (for IWOP) we introduced the pair $\# \ldots \#$ (for DOOT). The essence of the two techniques, namely *the commutativity of the creation and annihilation operators*, located inside the signs $\vdots \ldots \vdots$, respectively $\# \ldots \#$, is preserved [Popov, 2015b].

Like IWOP, DOOT is also based on the following simple rules:

*a*) Inside the symbol # # the order of the operators $\hat{\mathcal{A}}_-$ and $\hat{\mathcal{A}}_+$ can be permuted like commutable operators, but so that finally will result an operator function that depends only on the powers of normally ordered operator product $\hat{\mathcal{A}}_+\hat{\mathcal{A}}_-$, i.e.

$$\#\hat{\mathcal{A}}_-\hat{\mathcal{A}}_+\# = \#\hat{\mathcal{A}}_+\hat{\mathcal{A}}_-\# = \hat{\mathcal{A}}_+\hat{\mathcal{A}}_-\,, \qquad \#(\hat{\mathcal{A}}_-)^n(\hat{\mathcal{A}}_+)^n\# = \#(\hat{\mathcal{A}}_+)^n(\hat{\mathcal{A}}_-)^n\# = (\hat{\mathcal{A}}_+\hat{\mathcal{A}}_-)^n \tag{3.1}$$

*b*) A symbol # # inside another symbol # # can be deleted.

*c*) If the integration is convergent, a normally ordered product of operators can be integrated or differentiated, with respect to *c*-numbers, according to the usual rules. In addition, the *c*-numbers can be taken out from the symbol # #.

*d*) The projector $|0\rangle\langle 0|$ of the normalized vacuum state $|0\rangle$, in the frame of DOOT, has the following normal ordered form:

$$|0\rangle\langle 0| = \left[\sum_{n=0}^{\infty} \frac{\#\left(\hat{\mathcal{A}}_+\hat{\mathcal{A}}_-\right)^n\#}{{}_p\rho_n(n)}\right]^{-1} \tag{3.2}$$

As we will see below, these rules of the DOOT technique can be applied to both discrete and continuous spectrum systems. Regarding the DOOT technique, this provides a systematic way to achieve the normal ordering of creation and annihilation operators, while also allowing calculations with operators to be performed more easily. Therefore, this technique has found applications especially in quantum optics, in the context of constructing coherent states of various types.

### 3.2 Construction of generalized coherent states

To label the coherent states we will use the variable $z = |z|\exp(\mathrm{i}\varphi)$ with the ranges $0 < |z| \leq \infty$ and $0 \leq \varphi \leq 2\pi$. Generally, any nonlinear (or generalized coherent states) have the following mathematical structure:

$$|z>=\frac{1}{\sqrt{N\left(|z|^2\right)}}\sum_{n=0}^{n_{\max}\leq\infty}\frac{z^n}{\sqrt{{}_p\rho_n(n)}}|n> \quad , \qquad N\left(|z|^2\right)=\sum_{n=0}^{n_{\max}\leq\infty}\frac{1}{{}_p\rho_q(n)}\left(|z|^2\right)^n \tag{3.3}$$

where $\rho(n)$ are real and strictly positive numbers, called *structure constants*, because they determine the structure and character of the normalization function $N\left(|z|^2\right)$. This must be an positive analytical function, obtained from the normalization condition, $<z|z>=1$

We choose the following expression of the dimensionless energy eigenvalues

$${}_pe_q(n)\equiv n\,\frac{\prod_{j=1}^{q}\left[b_j+B_j(n-1)\right]}{\prod_{i=1}^{p}\left[a_i+A_i(n-1)\right]} \tag{3.4}$$

where $A_i$ and $B_j$ are positive, non-zero numbers. The advantage of this procedure is that it now also applies to quantum systems with a finite number of bound states (e.g., the Morse oscillator, spin systems, etc.). Then, the structure constant can have the following equivalent expressions

$$\begin{aligned}{}_p\rho_q(n)&=\prod_{s=1}^{n}{}_pe_q(s)=\prod_{s=1}^{n}\left[s\,\frac{\prod_{j=1}^{q}\left[b_j+B_j(s-1)\right]}{\prod_{i=1}^{p}\left[a_i+A_i(s-1)\right]}\right]=n!\,\frac{\prod_{j=1}^{q}\left(B_j\right)^n\left(\frac{b_j}{B_j}\right)_n}{\prod_{i=1}^{p}\left(A_i\right)^n\left(\frac{a_i}{A_i}\right)_n}=\\&=n!\frac{\prod_{j=1}^{q}\left(b_j\right)_{n,B_j}}{\prod_{i=1}^{p}\left(a_i\right)_{n,A_i}}=n!\frac{\prod_{i=1}^{p}\Gamma\left(a_i\right)}{\prod_{j=1}^{q}\Gamma\left(b_j\right)}\frac{\prod_{j=1}^{q}\Gamma\left(b_j+B_j\,n\right)}{\prod_{i=1}^{p}\Gamma\left(a_i+A_i\,n\right)}\end{aligned} \tag{3.5}$$

with the usual $\left(x\right)_n=\frac{\Gamma\left(x+n\right)}{\Gamma\left(x\right)}$, and the generalized Pochhammer symbols $\left(x\right)_{n,y}=y^n\left(\frac{x}{y}\right)_n$ .

The normalization function $N\left(|z|^2\right)$ of the generalized coherent states can be expressed through the Fox-Wright function ${}_p\Psi_q[\cdots]$, as [Juma, 2025]

$$N\left(|z|^2\right)=\sum_{n=0}^{n_{\max}\leq\infty}\frac{1}{{}_p\rho_q(n)}\left(|z|^2\right)^n=\frac{\prod_{j=1}^{q}\Gamma\left(b_j\right)}{\prod_{i=1}^{p}\Gamma\left(a_i\right)}\,{}_p\Psi_q\left[\begin{matrix}\left(a_1,A_1\right) & \left(a_2,A_2\right) & \ldots & \left(a_p,A_p\right)\\ \left(b_1,B_1\right) & \left(b_2,B_2\right) & \ldots & \left(b_q,B_q\right)\end{matrix};|z|^2\right] \tag{3.6}$$

The Fox-Wright function is defined as follows [Juma, 2025]:

$${}_p\Psi_q\left[\begin{matrix}\left(a_1,A_1\right) & \left(a_2,A_2\right) & \ldots & \left(a_p,A_p\right)\\ \left(b_1,B_1\right) & \left(b_2,B_2\right) & \ldots & \left(b_q,B_q\right)\end{matrix};|z|^2\right]=\sum_{n=0}^{n_{\max}\leq\infty}\frac{\prod_{i=1}^{p}\Gamma\left(a_i+A_i\,n\right)}{\prod_{j=1}^{q}\Gamma\left(b_j+B_j\,n\right)}\frac{\left(|z|^2\right)^n}{n!} \tag{3.7}$$

The normalization function can be expressed also through the generalized hypergeometric function due to their connection with the Fox-Wright function:

$$ {}_p\Psi_q\left[\begin{matrix}(a_1,A_1) & (a_2,A_2) & \ldots & (a_p,A_p)\\ (b_1,B_1) & (b_2,B_2) & \ldots & (b_q,B_q)\end{matrix};|z|^2\right]=\frac{\prod_{i=1}^{p}\Gamma(a_i)}{\prod_{j=1}^{q}\Gamma(b_j)}\,{}_pF_q\left(\left\{\left(\frac{a_i}{A_i}\right)_n\right\}_1^p;\left\{\left(\frac{b_j}{B_j}\right)_n\right\}_1^q;\frac{\prod_{i=1}^{p}A_i}{\prod_{j=1}^{q}B_j}|z|^2\right) \tag{3.8}$$

Also, the following limit will be useful to us further:

$$\lim_{\{A_i\}=\{B_j\}=1}\,{}_p\Psi_q\left[\begin{matrix}(a_1,A_1), & (a_2,A_2), & \ldots, & (a_p,A_p)\\ (b_1,B_1), & (b_2,B_2), & \ldots, & (b_q,B_q)\end{matrix};|z|^2\right]=$$
$$={}_p\Psi_q\left[\begin{matrix}(a_1,1), & (a_2,1), & \ldots, & (a_p,1)\\ (b_1,1), & (b_2,1), & \ldots, & (b_q,1)\end{matrix};|z|^2\right]={}_pF_q\left[\begin{matrix}a_1, & a_2, & \ldots, & a_p\\ b_1, & b_2, & \ldots, & b_q\end{matrix};|z|^2\right] \tag{3.9}$$

In this sense, it can be said that the Fox-Wright function ${}_p\Psi_q$ is actually a usual generalization of the generalized hypergeometric function ${}_pF_q$. We point out that if the maximum number of bound states is infinite, $n_{\max}=\infty$, the above function is a generalized series, while if this maximum number is finite, $n_{\max}<\infty$ we obtain the generalized polynomials of degree $n_{\max}$.

In the following, in order to simplify the writing of mathematical relations, we will use the following simplifying notations: for the Fox-Wright function, ${}_p\Psi_q(x)$, or ${}_p\Psi_q\left(\begin{matrix}(\mathrm{a},\mathrm{A})\\(\mathrm{b},\mathrm{B})\end{matrix}\middle| x\right)$, as well as for the generalized hypergeometric function, ${}_pF_q(x)$, or, where confusion could arise, we will use the full notation ${}_pF_q\left(\left\{\left(\frac{a_i}{A_i}\right)_n\right\}_1^p;\left\{\left(\frac{b_j}{B_j}\right)_n\right\}_1^q;\frac{\prod_{i=1}^{p}A_i}{\prod_{j=1}^{q}B_j}x\right)$.

In conclusion, the most complete generalized coherent states (which here are states that depend on a single complex variable, although coherent states that depend on multiple variables can also be defined !) are those whose normalization function $N(|z|^2)$ is proportional with the Fox-Wright function, or, equivalently, the normalization function $N(|z|^2)$ is a generalized hypergeometric function ${}_pF_q\left(\left\{\left(\frac{a_i}{A_i}\right)_n\right\}_1^p ; \left\{\left(\frac{b_j}{B_j}\right)_n\right\}_1^q ; \frac{\prod_{i=1}^{p} A_i}{\prod_{j=1}^{q} B_j} \mid z\mid^2\right)$, with the positive integers $p$ and $q$ as parameters, and also $\{a_i\}_1^p = a_1, a_2, ..., a_p$, $\{A_i\}_1^p = A_1, A_2, ..., A_p$, $\{b_j\}_1^q = b_1, b_2, ..., b_q$, $\{B_j\}_1^q = B_1, B_2, ..., B_q$ being real numbers. The reason is simple: the generalized hypergeometric function is one of the most general functions of a single variable, from which, by specifying the indices and constants it contains, almost all special functions used in the analysis of physical and technical phenomena can be deduced.

Therefore, for the most general expression of a normalization function for the generalized coherent states we will use the generalized hypergeometric function, which has the expression

$$N\left(\mid z\mid^2\right) = {}_pF_q\left(\left\{\left(\frac{a_i}{A_i}\right)_n\right\}_1^p ; \left\{\left(\frac{b_j}{B_j}\right)_n\right\}_1^q ; \frac{\prod_{i=1}^{p} A_i}{\prod_{j=1}^{q} B_j} \mid z\mid^2\right) =$$

$$= \sum_{n=0}^{n_{\max}\le\infty} \frac{\prod_{i=1}^{p}\left(\frac{a_i}{A_i}\right)_n}{\prod_{j=1}^{q}\left(\frac{b_j}{B_j}\right)_n} \frac{\left(\frac{\prod_{i=1}^{p} A_i}{\prod_{j=1}^{q} B_j} \mid z\mid^2\right)^n}{n!} \equiv \sum_{n=0}^{n_{\max}\le\infty} \frac{1}{{}_p\rho_q(n)} \frac{\left(\mid z\mid^2\right)^n}{n!} \tag{3.10}$$

An additional reason is the fact that many generalized hypergeometric functions are expressible by more general functions or equivalent functions (see the *wolfram functions site*, web page https://functions.wolfram.com/HypergeometricFunctions/ ).

Consequently, the more generalized set of coherent states have the following expression:

$$\mid z\,\rangle = \frac{1}{\sqrt{{}_pF_q\left(\left\{\left(\frac{a_i}{A_i}\right)_n\right\}_1^p ; \left\{\left(\frac{b_j}{B_j}\right)_n\right\}_1^q ; \frac{\prod_{i=1}^{p} A_i}{\prod_{j=1}^{q} B_j} \mid z\mid^2\right)}} \sum_{n=0}^{n_{\max}\le\infty} \sqrt{\frac{\prod_{i=1}^{p}\left(a_i\right)_{n,A_i}}{\prod_{j=1}^{q}\left(b_j\right)_{n,B_j}}} \frac{z^n}{\sqrt{n!}} \mid n\,\rangle \tag{3.11}$$

or, in the abbreviated version

$$| z >= \frac{1}{\sqrt{{}_pF_n\left(| z |^2\right)}} \sum_{n=0}^{n_{\max} \leq \infty} \frac{z^n}{\sqrt{{}_p\rho_n(n)}} | n > \qquad (3.12)$$

By suitable particularizing the sets of indices $p$ and $q$, as well as the sequences of numbers $\{a_i , A_i\}_1^p$ and $\{b_j , B_j\}_1^q$, one can obtain all types of coherent states associated with different quantum systems, and therefore which have physical meaning. Generally, the two sets of parameters, in numerator and denominator, must be nonzero numbers. The series terminates if either $\{a_i\}_1^p$ is a negative integer, in which case the function reduces to a hypergeometric polynomial. In order to have the above normalization function $N\left(| z |^2\right)$ the nonlinearity function must have the following expression

$$f(\hat{\mathcal{N}}) = \sqrt{\frac{\prod_{j=1}^{q}\left[b_j + B_j\left(\hat{\mathcal{N}} - 1\right)\right]}{\prod_{i=1}^{p}\left[a_i + A_i\left(\hat{\mathcal{N}} - 1\right)\right]}} \quad , \quad f(\hat{\mathcal{N}}) | n >= \sqrt{\frac{\prod_{j=1}^{q}\left[b_j + B_j(n-1)\right]}{\prod_{i=1}^{p}\left[a_i + A_i(n-1)\right]}} | n > \qquad (3.13)$$

***Definition I*: Generalized coherent states defined as eigenvectors of the annihilation operator**. In the spirit of the ideas of Glauber [Glauber, 1963] and, a few years later, of Barut and Girardello [Barut, 1971], for the construction of generalized coherent states we will use the definition according to which they are eigenvectors of the annihilation operator $\hat{\mathcal{A}}_- = \hat{a}\, f\left(\hat{\mathcal{N}}\right)$, i.e. $\hat{\mathcal{A}}_- | z >= z | z >$.

***Notes***: Generally, the system's Hamiltonian depend of some parameters. Consequently, apart from the variable $z$, the coherent states will also depend on some parameters (for example, the orbital $J$ or spin $S$ quantum numbers, etc.). However, in order not to complicate writing formulas, we will generally not mention these parameters, but only where strictly necessary to avoid confusion.

Like any other state vector, the vectors representing the coherent states can be expanded as a linear combination of the basic Fock vectors. This can be easily obtained, taking into consideration both the definition of coherent states and the result of the action of the nonlinear annihilation operator on the Fock vectors. So, this expansion and their dual are

$$| z >= \frac{1}{\sqrt{{}_pF_q\left(| z |^2\right)}} \sum_{n=0}^{n_{\max} \leq \infty} \frac{z^n}{\sqrt{{}_p\rho_q(n)}} | n > \quad , \quad < z | = \frac{1}{\sqrt{{}_pF_q\left(| z |^2\right)}} \sum_{n=0}^{n_{\max} \leq \infty} \frac{\left(z^*\right)^n}{\sqrt{{}_p\rho_q(n)}} < n | \qquad (3.14)$$

The convergence radius $R_c$ , if $A_i \neq 0$ is

$$R_c = \lim_{n\to\infty} \frac{{}_p\rho_q(n)}{{}_p\rho_q(n+1)} = \lim_{n\to\infty} n^{p-q-1} \frac{\prod_{j=1}^{q} B_j}{\prod_{i=1}^{p} A_i} = \begin{cases} \infty, & \text{if} \quad p-q-1>0 \quad \text{infinite convergence radius} \\ \frac{\prod_{j=1}^{q} B_j}{\prod_{i=1}^{p} A_i} \neq 0 & \text{if} \quad p-q-1=0 \quad \text{finite convergence radius} \\ 0 & \text{if} \quad p-q-1<0 \quad \text{coherent states do not exist} \end{cases}$$

According to Klauder's considerations, later called "Klauder's minimal prescriptions", for a vector $|z>$ to be considered a coherent state, it must meet the following fundamental properties [Gazeau, 1999], [Roy, 2006]:

1.) *Continuity in the labeling variable*: $z' \to z \quad \Leftrightarrow \quad \| |z'> - |z> \| \to 0$.

2.) *Normalization and non-orthogonality*:

$$< z | z' > = \frac{{}_pF_q\left(z^* z'\right)}{\sqrt{{}_pF_q\left(|z|^2\right)}\sqrt{{}_pF_q\left(|z'|^2\right)}} = \begin{cases} 1 & \text{for} \quad z = z' \quad \text{normalization condition} \\ \neq 0 & \text{for} \quad z \neq z' \quad \text{non orthogonality condition} \end{cases}$$

3.) *Resolution of the unity operator*: $\int d\mu(z)\, |z><z| = I$, where the integration measure $d\mu(z) = \frac{d^2 z}{\pi} h\left(|z|^2\right) = \frac{d\varphi}{\pi}\, d\left(|z|^2\right) h\left(|z|^2\right)$ with the weight function $h\left(|z|^2\right)$ which must be a positive and unique function.

3.) *Temporal stability*: $e^{-i\hat{\mathcal{H}} t}\, |z> = e^{-i E_0 t}\, |z(t)>$ , $E_n = E_0 + \hbar\omega n$ , $z(t) = e^{-i\hbar\omega t} z$ , i.e. any coherent state remains coherent during the time: $|z(t_0 = 0)> \to |z(t)>$ .

4.) *Action identity*: $H(z) \equiv < z | \hat{\mathcal{H}} | z > = \omega |z|^2$ , where $E_0 = 0$ , $\hbar = 1$ . The name "action identity" comes from the fact that the parameter $|z|^2$ can be interpreted as the classical action variable conjugate with the angle variable $\omega$ .

But, the last two conditions are fulfilled only by the coherent states corresponding to systems with linear energy spectrum $E_n \cong \hbar\omega\, n$ .

From the definition of CSs it follows that is valid the quantization scheme

| ***Classical state*** $\{q,\ p\}$ | $\leftrightarrow$ | $z \to \ |z><z|$ ***Quantum state*** |
|---|---|---|
| *Phase space* $\mathcal{X}$ | $\leftrightarrow$ | *Hilbert space* $\mathcal{H}$ |

as imagined Schrodinger in 1926. From this point of view, the CSs provide a close connection between classical and quantum mechanics.

Let us now find the expression of the integration measure that appears in the decomposition relation of the unit operator. Substituting the expressions of the coherent states, we will have:

$$I = \int d\mu(z)\, |z><z| = \sum_{n=0}^{\infty} \frac{|n>}{\sqrt{{}_p\rho_q(n)}} \sum_{m=0}^{\infty} \frac{<m|}{\sqrt{{}_p\rho_q(m)}} \int_0^{\infty} d\left(|z|^2\right) \frac{h\left(|z|^2\right)}{{}_pF_q\left(|z|^2\right)} \int_0^{2\pi} \frac{d\varphi}{2\pi} z^n \left(z^*\right)^m \qquad (3.15)$$

In the end the closure (or completeness) relation of the Fock vectors must be satisfied $\sum_{n=0}^{\infty} | n >< n | = 1$. The angular integral is $\left(| z |^2\right)^n \delta_{nm}$, so we will have

$$I = \sum_{n=0}^{\infty} \frac{| n >< n |}{{}_p\rho_q(n)} \int_0^{\infty} d\left(| z |^2\right) \frac{h\left(| z |^2\right)}{{}_pF_q\left(| z |^2\right)} \left(| z |^2\right)^n \tag{3.16}$$

Since the unknown function is $\tilde{h}\left(| z |^2\right) \equiv \frac{h\left(| z |^2\right)}{{}_pF_q\left(| z |^2\right)}$, obviously we must have the equality

$$\int_0^{\infty} d\left(| z |^2\right) \tilde{h}\left(| z |^2\right)\left(| z |^2\right)^n = {}_p\rho_q(n) \tag{3.17}$$

From a mathematical point of view, such an equation is called the moment problem. In problems related to coherent states, two types of moments arise: the Stieltjes moment problem, for the entire positive axis $(0\,,\,+\infty)$, and the Hausdorff moment problem for a bounded interval, $(0\,,\,1)$ [Schmüdgen, 2017].

Because the structure constant can also be written in the manner

$${}_p\rho_q(n) = n! \frac{\prod_{j=1}^{q} \left(b_j\right)_{n,B_j}}{\prod_{i=1}^{p} \left(a_i\right)_{n,A_i}} = \Gamma\left(\frac{a}{A} \,/\, \frac{b}{B}\right) \left(\frac{\prod_{j=1}^{q} B_j}{\prod_{i=1}^{p} A_i}\right)^n \Gamma(n+1) \frac{\prod_{j=1}^{q} \Gamma\left(\frac{b_j}{B_j} + n\right)}{\prod_{i=1}^{p} \Gamma\left(\frac{a_i}{A_i} + n\right)}\,, \quad \Gamma\left(\frac{a}{A} \,/\, \frac{b}{B}\right) \equiv \frac{\prod_{i=1}^{p} \Gamma\left(\frac{a_i}{A_i}\right)}{\prod_{j=1}^{q} \Gamma\left(\frac{b_j}{B_j}\right)} \tag{3.18}$$

we will adapt for our case the classical formula of a generalized integral for a single Meijer G function [Mathai, 1973]:

$$\int_0^{\infty} dx\, x^{s-1}\, G_{p,q}^{m,n}\left(\omega x \left| \begin{array}{ll} \{\alpha_i\}_1^n\,; & \{\alpha_i\}_{n+1}^p \\ \{\beta_j\}_1^m\,; & \{\beta_j\}_{m+1}^q \end{array} \right.\right) = \frac{1}{\omega^s} \frac{\prod_{j=1}^{m} \Gamma\left(\beta_j + s\right) \prod_{i=1}^{n} \Gamma\left(1 - \alpha_i - s\right)}{\prod_{j=m+1}^{q} \Gamma\left(1 - \beta_j - s\right) \prod_{i=n+1}^{p} \Gamma\left(\alpha_i + s\right)} \tag{3.19}$$

It is observed that in the "classic" formula (in which $\{A_i\}_1^p = \{B_j\}_1^q = 1$), now, in the new situation, when $\{A_i\}_1^p \neq \{B_j\}_1^q \neq 1$, the following replacements (or "renormalizations" of the sets of numbers) must be made:

$$\Gamma(a/b) \leftrightarrow \Gamma\left(\frac{a}{A} \,/\, \frac{b}{B}\right)\,, \quad \{a_i\}_1^p \leftrightarrow \left\{\frac{a_i}{A_i}\right\}_1^p\,, \quad \{b_j\}_1^q \leftrightarrow \left\{\frac{b_j}{B_j}\right\}_1^q \tag{3.20}$$

It is easy to see that the unknown function sought is proportional to Meijer's G function and finally the expression for the integration measure will be

$$d\mu(z)=\Gamma\left(\frac{a}{A}\,/\,\frac{b}{B}\right)\frac{\prod_{j=1}^{q}B_j}{\prod_{i=1}^{p}A_i}\frac{d\varphi}{2\pi}d\left(|z|^2\right)G_{p,q+1}^{q+1,0}\left(\frac{\prod_{i=1}^{p}A_i}{\prod_{j=1}^{q}B_j}|z|^2\left|\begin{array}{cc} /; & \left\{\frac{a_i}{A_i}-1\right\}_1^p \\ 0,\left\{\frac{b_j}{B_j}-1\right\}_1^q; & / \end{array}\right.\right)\times$$
$$\times{}_pF_q\left(\left\{\left(\frac{a_i}{A_i}\right)_n\right\}_1^p;\left\{\left(\frac{b_j}{B_j}\right)_n\right\}_1^q;\frac{A_i}{B_j}|z|^2\right) \tag{3.21}$$

The expectation value of operator $\hat{O}$ in the generalized coherent states representation is

$$<z|\hat{O}|z>\equiv<\hat{O}>_z=\frac{1}{{}_pF_q\left(|z|^2\right)}\sum_{n',n=0}^{n_{\max}\leq\infty}\frac{(z^*)^{n'}z^n}{\sqrt{{}_p\rho_q(n')\;{}_p\rho_q(n)}}<n'|\hat{O}|n> \tag{3.22}$$

The expected value of the normal ordered product $\#\hat{\mathcal{A}}_+\hat{\mathcal{A}}_-\#$ in the Fock basis is

$$<n|\#\hat{\mathcal{A}}_+\hat{\mathcal{A}}_-\#|n>=e(n) \tag{3.23}$$

Consequently, in the generalized coherent states representation, the expected value is

$$<\#\hat{\mathcal{A}}_+\hat{\mathcal{A}}_-\#>_z=\frac{1}{{}_pF_q\left(|z|^2\right)}\sum_{n=0}^{n_{\max}\leq\infty}\frac{\left(|z|^2\right)^n}{{}_p\rho_q(n)}e(n)=\frac{1}{{}_pF_q\left(|z|^2\right)}e\left(\vartheta_{|z|^2}\right){}_pF_q\left(|z|^2\right) \tag{3.24}$$

On the other hand, from the eigenvalue equation of the normal ordered operator’s product $\hat{\mathcal{A}}_+\hat{\mathcal{A}}_-$ and using the DOOT expression of the vacuum projector, we can writte also

$$\hat{\mathcal{A}}_+\hat{\mathcal{A}}_-=\sum_{n=0}^{\infty}e(n)|n><n|=\frac{1}{\#{}_pF_q\left(\hat{\mathcal{A}}_+\hat{\mathcal{A}}_-\right)\#}\sum_{n=0}^{\infty}{}_pe_q(n)\frac{\#\left(\hat{\mathcal{A}}_+\hat{\mathcal{A}}_-\right)^n\#}{{}_p\rho_n(n)} \tag{3.25}$$

Above we use the theta (or Euler) operator $\vartheta_x\equiv x\frac{\partial}{\partial x}$, which satisfies the equalities

$$nx^n=x\frac{\partial}{\partial x}x^n\quad,\quad(\vartheta_x+c)x^nx^n=(n+c)x^n\quad,\quad e(n)x^n=e\left(\vartheta_x\right)x^n \tag{3.26}$$

This means that in expressions of this type the following substitution can be made $n\leftrightarrow x\frac{\partial}{\partial x}=\vartheta_x$, or $n\leftrightarrow\hat{\mathcal{A}}_+\hat{\mathcal{A}}_-\frac{\partial}{\partial\hat{\mathcal{A}}_+\hat{\mathcal{A}}_-}=\vartheta_{\hat{\mathcal{A}}_+\hat{\mathcal{A}}_-}$, and then the above expression becomes

$$\hat{\mathcal{A}}_+\hat{\mathcal{A}}_-=\frac{1}{\#{}_pF_q\left(\hat{\mathcal{A}}_+\hat{\mathcal{A}}_-\right)\#}e\left(\vartheta_{\hat{\mathcal{A}}_+\hat{\mathcal{A}}_-}\right)\#{}_pF_q\left(\hat{\mathcal{A}}_+\hat{\mathcal{A}}_-\right)\# \tag{3.27}$$

or, equivalently

$$\left[e\left(\vartheta_{\hat{\mathcal{A}}_+\hat{\mathcal{A}}_-}\right)-\hat{\mathcal{A}}_+\hat{\mathcal{A}}_-\right]\#{}_pF_q\left(\hat{\mathcal{A}}_+\hat{\mathcal{A}}_-\right)\#=0 \tag{3.28}$$

If we pass to the variable $\hat{\mathcal{A}}_+\hat{\mathcal{A}}_-\to|z|^2$, we will obtain the equation

$$\left[e\left(\vartheta_{|z|^2}\right)-|z|^2\right]{}_pF_q\left(|z|^2\right)=0 \tag{3.29}$$

which shows that the eigenvalue of the operator $e\left(\vartheta_{|z|^2}\right)$, corresponding to the eigenfunction ${}_pF_q\left(|z|^2\right)$, is precisely the square of the modulus of label variable $|z|^2$ of the coherent states.

If we write this equality like this

$$e\left(\vartheta_{|z|^2}\right)\ {}_pF_q\left(|z|^2\right) = <\#\hat{\mathcal{A}}_+\hat{\mathcal{A}}_-\#>_z\ {}_pF_q\left(|z|^2\right) \tag{3.30}$$

we can say that the generalized hypergeometric function ${}_pF_q\left(|z|^2\right)$ is an eigenfunction of the structure constant theta operator $e\left(\vartheta_{|z|^2}\right)$. This is obtained by replacing the quantum number $n$ with the theta operator corresponding to the variable $|z|^2$, that is $n \leftrightarrow \vartheta_{|z|^2} \equiv |z|^2 \dfrac{\partial}{\partial |z|^2}$. Then, the expected average value, in the representation of coherent states, $<\#\hat{\mathcal{A}}_+\hat{\mathcal{A}}_-\#>_z$, is just the corresponding eigenvalue.

$$\left[e\left(\vartheta_{|z|^2}\right) - <\#\hat{\mathcal{A}}_+\hat{\mathcal{A}}_-\#>_z\right]\ {}_pF_q\left(|z|^2\right) = 0 \tag{3.31}$$

Comparing this result with those obtained above, we obtain that the expected average value of the ordered product of the operators is

$$<\#\hat{\mathcal{A}}_+\hat{\mathcal{A}}_-\#>_z = |z|^2 \tag{3.32}$$

Therefore, we obtain a new differential equation for generalized hypergeometric function, in addition to the one that appears in the literature [Letessier, 1994]:

$$\left[{}_pe_q\left(\vartheta_{|z|^2}\right) - |z|^2\right]\ {}_pF_q\left(|z|^2\right) = 0 \tag{3.33}$$

where the following correspondence exists

$${}_pe_q(n) \equiv n\,\frac{\prod_{j=1}^{q}\left[b_j + B_j(n-1)\right]}{\prod_{i=1}^{p}\left[a_i + A_i(n-1)\right]} \quad \leftrightarrow \quad {}_pe_q\left(\vartheta_{|z|^2}\right) \equiv \frac{\vartheta_{|z|^2}\prod_{j=1}^{q}\left[b_j + B_j(\vartheta_{|z|^2}-1)\right]}{\prod_{i=1}^{p}\left[a_i + A_i(\vartheta_{|z|^2}-1)\right]} \tag{3.34}$$

In an equivalent manner, this can be written as

$$\left[\vartheta_{|z|^2}\prod_{j=1}^{q}\left[b_j + B_j(\vartheta_{|z|^2}-1)\right] - |z|^2\prod_{i=1}^{p}\left[a_i + A_i(\vartheta_{|z|^2}-1)\right]\right]\ {}_pF_q\left(|z|^2\right) = 0 \tag{3.35}$$

This is another differential equation satisfied by generalized hypergeometric functions ${}_pF_q\left(|z|^2\right)$.

The same relation is obtained if the scalar product of the definition relation of coherent states of the Barut-Girardello type, with its adjoint relation

$$<\#\hat{\mathcal{A}}_+\hat{\mathcal{A}}_-\#>_z = \left(<z|\,\hat{\mathcal{A}}_+\right)\left(\hat{\mathcal{A}}_-\,|z>_z\right) = z^* z = |z|^2 \tag{3.36}$$

Similarly, we have, for a integer power $m$

$$<\#\left(\hat{\mathcal{A}}_+\hat{\mathcal{A}}_-\right)^m\#>_z = \frac{1}{{}_pF_q\left(|z|^2\right)}\left[e\left(\vartheta_{|z|^2}\right)\right]^m\ {}_pF_q\left(|z|^2\right) \tag{3.37}$$

$$\left[e\left(\vartheta_{|z|^2}\right)\right]^m\ {}_pF_q\left(|z|^2\right) = <\#\left(\hat{\mathcal{A}}_+\hat{\mathcal{A}}_-\right)^m\#>_z\ {}_pF_q\left(|z|^2\right) \tag{3.38}$$

Consequently, the expected average value of any function that depends on the normal ordered product of $\#\hat{\mathcal{A}}_+\hat{\mathcal{A}}_-\#$ will be obtained by replacing the normal ordered product of the operators with the $e\left(\vartheta_{|z|^2}\right)$, i.e. $<\#\hat{\mathcal{A}}_+\hat{\mathcal{A}}_-\#>_z \leftrightarrow e\left(\vartheta_{|z|^2}\right)$ :

$$< \mathcal{F}\left(\#\hat{\mathcal{A}}_+\hat{\mathcal{A}}_-\#\right)>_z = \sum_{n=0}^{\infty}\frac{1}{n!}\mathcal{F}^{(n)}(0) < \left(\#\hat{\mathcal{A}}_+\hat{\mathcal{A}}_-\#\right)^n >_z = \sum_{n=0}^{\infty}\frac{1}{n!}\mathcal{F}^{(n)}(0)\left(|z|^2\right)^n = \mathcal{F}\left(|z|^2\right) \qquad (3.39)$$

A useful example is is the integer power case of the particle number operator $\hat{O} = \hat{\mathcal{N}}^s$, where $s = 1,2,...$ and $\hat{\mathcal{N}}$ is the particle number operator, i. e. $\hat{\mathcal{N}}|n>= n|n>$, then the expectation value can be

$$< \hat{\mathcal{N}}^s >_z = \frac{1}{{}_pF_q\left(|z|^2\right)}\left[|z|^2 \frac{d}{d(|z|^2)}\right]^s {}_pF_q\left(|z|^2\right) = \frac{1}{{}_pF_q\left(|z|^2\right)}\sum_{n=0}^{\infty}\frac{\left(|z|^2\right)^n}{{}_p\rho_q(n)}n^s \qquad (3.40)$$

To compare the statistics of different generalized coherent states we compute the Mandel parameter, defined as follows [Mandel, 1994], which in our case is

$$Q_{|z|} = \frac{< \hat{\mathcal{N}}^2 >_z - \left(< \hat{\mathcal{N}} >_z\right)^2}{< \hat{\mathcal{N}} >_z} - 1 = \frac{\sum_{n=0}^{n_{\max}\le\infty}\frac{\left(|z|^2\right)^n}{{}_p\rho_q(n)}n^2}{\sum_{n=0}^{n_{\max}\le\infty}\frac{\left(|z|^2\right)^n}{{}_p\rho_q(n)}n} - \frac{\sum_{n=0}^{n_{\max}\le\infty}\frac{\left(|z|^2\right)^n}{{}_p\rho_q(n)}n}{\sum_{n=0}^{n_{\max}\le\infty}\frac{\left(|z|^2\right)^n}{{}_p\rho_q(n)}} - 1 \qquad (3.41)$$

Depending on the values of the Mandel parameter, the statistics to which coherent states are subject are the following:

$$Q_{|z|} = \begin{cases} < 0 & \text{sub - Poissonian statistics (anti - bunched light, quantum, e.g. single photon sources)} \\ = 0 & \text{Poissonian statistics (coherent light, classical limits, e.g. ideal laser)} \\ > 0 & \text{super - Poissonian statistics (bunched light, classical states, e.g. thermal light)} \end{cases}$$

In our case, because $\sum_{n=0}^{n_{\max}\le\infty}\frac{\left(|z|^2\right)^n}{{}_p\rho_q(n)}n^2 > \sum_{n=0}^{n_{\max}\le\infty}\frac{\left(|z|^2\right)^n}{{}_p\rho_q(n)}n$ , and substituting into the expression of Mandel's parameter, we obtain:

$$Q_{|z|} < 1 - \frac{\sum_{n=0}^{n_{\max}\le\infty}\frac{\left(|z|^2\right)^n}{{}_p\rho_q(n)}n}{\sum_{n=0}^{n_{\max}\le\infty}\frac{\left(|z|^2\right)^n}{{}_p\rho_q(n)}} - 1 < 0 \qquad (3.42)$$

This expression is negative because ${}_p\rho_q(n) = \sum_{j=1}^{n} e(j) > 0$, as well as $E(n) = \hbar\,\omega\, e(n) > 0$ , so that, for generalized coherent states, the Mandel parameter is negative, $Q_{|z|} < 0$, therefore they obey a sub-Poissonian statistics.

Generally, because the hypergeometric functions are present, the Mandel parameter must be evaluated numerically for different variable values in order to establish the sign of $Q_{|z|}$ will be. Only in a few particular cases, the Mandel parameter can be evaluated analytically.

The generalized probability density of finding $n$ photons in the coherent state $|z>$ is

$$\mathcal{P}_n\left(|z|^2\right)=|<z|n>|^2=\frac{1}{{}_pF_q\left(|z|^2\right)}\frac{\left(|z|^2\right)^n}{{}_p\rho_q(n)}\Leftrightarrow\mathcal{P}_n^{(\text{Poisson})}\left(|z|^2\right)=e^{-|z|^2}\frac{\left(|z|^2\right)^n}{n!}\tag{3.43}$$

which lead to the following kind of didtributions:

$$\mathcal{P}_n\left(|z|^2\right)\begin{cases}< \ \mathcal{P}_n^{(\text{Poisson})}\left(|z|^2\right) & \text{sub Poissonian}\\ = \ \mathcal{P}_n^{(\text{Poisson})}\left(|z|^2\right) & \text{Poissonian}\\ > \ \mathcal{P}_n^{(\text{Poisson})}\left(|z|^2\right) & \text{super Poissonian}\end{cases}\tag{3.44}$$

Taking into account the definitions of the generalized hypergeometric function, as well as the structure constant of coherent states, we can write the following limits:

$$\lim_{\substack{p=q\\ \{a_i\}=\{b_j\}\\ \{A_i\}=\{B_j\}=1}} {}_p\rho_q(n)=n! \quad , \quad \lim_{\substack{p=q\\ \{a_i\}=\{b_j\}\\ \{A_i\}=\{B_j\}=1}} {}_pF_q\left(|z|^2\right)=\sum_{n=0}^{n_{\max}\leq\infty}\frac{\left(|z|^2\right)^n}{\lim\limits_{\substack{p=q\\ \{a_i\}=\{b_j\}\\ \{A_i\}=\{B_j\}=1}} {}_p\rho_q(n)}=\sum_{n=0}^{\infty}\frac{\left(|z|^2\right)^n}{n!}=e^{|z|^2}\tag{3.45}$$

which shows that the generalized probability distributions tend towards the Poisson distribution.

Using the pair of generalized operators $\hat{\mathcal{A}}_+$ and $\hat{\mathcal{A}}_-$, let's construct a pair of generalized quadrature operators, i.e. the dimensionless coordinate $\hat{\mathcal{Q}}$ and the momentum $\hat{\mathcal{P}}$, which are the Hermitian operators

$$\hat{\mathcal{Q}}=\frac{1}{\sqrt{2}}\left(\hat{\mathcal{A}}_+ + \hat{\mathcal{A}}_-\right) \quad , \quad \hat{\mathcal{P}}=\mathrm{i}\frac{1}{\sqrt{2}}\left(\hat{\mathcal{A}}_+ - \hat{\mathcal{A}}_-\right)\tag{3.46}$$

and reciprocally

$$\hat{\mathcal{A}}_-=\frac{1}{\sqrt{2}}\left(\hat{\mathcal{Q}}+\mathrm{i}\,\hat{\mathcal{P}}\right) \quad , \quad \hat{\mathcal{A}}_+=\frac{1}{\sqrt{2}}\left(\hat{\mathcal{Q}}-\mathrm{i}\,\hat{\mathcal{P}}\right)\tag{3.47}$$

Our aim is to calculate the different expected values, in the representation of the coherent states, previously defined, with the final goal of writing the uncertainty relations for the coordinate - momentum pair, in the representation of generalized coherent states.

$$\begin{pmatrix}<\hat{\mathcal{A}}_->_z\\ <\hat{\mathcal{A}}_+>_z\end{pmatrix}=\begin{pmatrix}z\\ z^*\end{pmatrix} \quad , \quad \begin{pmatrix}<\left(\hat{\mathcal{A}}_-\right)^2>_z\\ <\left(\hat{\mathcal{A}}_+\right)^2>_z\end{pmatrix}=\begin{pmatrix}z^2\\ \left(z^*\right)^2\end{pmatrix}\tag{3.48}$$

$$<n|\left(\hat{\mathcal{A}}_+\hat{\mathcal{A}}_-\right)|n>=e(n) \quad , \quad <n|\left(\hat{\mathcal{A}}_-\hat{\mathcal{A}}_+\right)|n>=e(n+1)\tag{3.49}$$

$$<\left(\hat{\mathcal{A}}_+\hat{\mathcal{A}}_-\right)>_z=|z|^2 \quad , \quad <\left(\hat{\mathcal{A}}_-\hat{\mathcal{A}}_+\right)>_z=\frac{1}{{}_pF_q\left(|z|^2\right)}\sum_{n=0}^{n_{\max}\leq\infty}\frac{\left(|z|^2\right)^n}{{}_p\rho_q(n)}\,e(n+1)\tag{3.50}$$

Therefore, the expected average value for the commutator of the two operators will be

$$<\left[\hat{\mathcal{A}}_- \, , \, \hat{\mathcal{A}}_+\right]>_z=\frac{1}{{}_pF_q\left(|z|^2\right)}\sum_{n=0}^{n_{\max}\leq\infty}\frac{\left(|z|^2\right)^n}{{}_p\rho_q(n)}\left[e(n+1)-e(n)\right]\tag{3.51}$$

The right parenthesis on the right-hand side of this equality represents precisely the discrete derivatives (also called the finite difference) of the function $e(n)$ that depends on the discrete variable $n$:

$$\frac{\Delta}{\Delta n} e(n) = e(n+1) - e(n) \tag{3.52}$$

From another point of view, the function $e(n)$ can also be viewed as the eigenvalue of the operator $e(\hat{\mathcal{N}})$ which depends on the particle number operator $\hat{\mathcal{N}}$:

$$e(\hat{\mathcal{N}})| n > = e(n) | n > \tag{3.53}$$

so that discrete derivatives will be written operationally as follows:

$$\frac{\Delta}{\Delta \hat{\mathcal{N}}} e(\hat{\mathcal{N}}) = e(\hat{\mathcal{N}}+1) - e(\hat{\mathcal{N}}) = (\hat{\mathcal{T}}_1 - 1) e(\hat{\mathcal{N}}) \tag{3.54}$$

where $\hat{\mathcal{T}}_1$ is the translation operator, which has the property $\hat{\mathcal{T}}_k \, e(\hat{\mathcal{N}}) = e(\hat{\mathcal{N}} + k)$.

In this way the expected average value of the commutation relation of the generalized annihilation and creation operators will be written as

$$< [\hat{\mathcal{A}}_- \, , \, \hat{\mathcal{A}}_+] >_z = \frac{1}{{}_pF_q(| z |^2)} \sum_{n=0}^{n_{\max} \leq \infty} \frac{(| z |^2)^n}{{}_p\rho_q(n)} (\hat{\mathcal{T}}_1 - 1) e(n) = (\hat{\mathcal{T}}_1 - 1) < e(\hat{\mathcal{N}}) >_z \tag{3.55}$$

This equality will be useful in expressing the generalized uncertainty relation for a pair of noncommutative operators (i.e. with nonzero commutator):

$$(\Delta \hat{\mathcal{A}}_-)_z^2 (\Delta \hat{\mathcal{A}}_+)_z^2 \geq \frac{1}{4} < z | [\hat{\mathcal{A}}_- \, , \, \hat{\mathcal{A}}_+] | z >_z|^2 \tag{3.56}$$

$$(\Delta \hat{\mathcal{A}}_-)_z (\Delta \hat{\mathcal{A}}_+)_z \geq \frac{1}{2} \sqrt{(\hat{\mathcal{T}}_1 - 1) | < e(\hat{\mathcal{N}}) >_z |} \tag{3.57}$$

Since the pair of annihilation / creation operators does not commute, they satisfy the above relation. Therefore, generalized coherent states do not saturate the uncertainty relation.

But, at the harmonic limit, i. e. for $f(\hat{\mathcal{N}}) \to 1$, we will have $e(n) = n$ we obtain

$$\lim_{f(\hat{\mathcal{N}}) \to 1} (\Delta \hat{\mathcal{A}}_-)_z (\Delta \hat{\mathcal{A}}_+)_z = (\Delta \hat{\boldsymbol{a}})_z (\Delta \hat{\boldsymbol{a}}_+)_z = \frac{1}{2} \tag{3.58}$$

so we will expected, and obtain the canonical commutation relations, characteristic for HO-1D, in which the uncertainty product is minimized. This means that the canonical coherent states saturate the uncertainty relation.

***Definition II:*** **Generalized coherent states defined as the result of the application of the creation operator on the vacuum state.** We will consider the definition in which the generalized displacement operator intervenes which, for the case of the pair of canonical operators, led to the expression of coherent states of the Klauder-Perelomov type [Klauder, 1963], [Perelomov, 1972], [Roy, 2000]. This definition consists of applying the exponential displacement operator $\exp(z \boldsymbol{a}^+ - z^* \boldsymbol{a})$, to the vacuum state $| 0 >$ simultaneously taking into account that the action of the annihilation operator on these states is null $\hat{\boldsymbol{a}} \, | 0 > = 0$.

To highlight another definition of generalized coherent states we will use an idea introduced in Mojaveri and Dehghani's paper [Mojaveri, 2013], in which the generalized displacement operator is expressed by a generalized hypergeometric function that includes, as an argument, the generalized creation operator, taking into account the relation $\hat{\mathcal{A}}_- \,|\, 0>=0$.

According to DOOT rules, we will have [Popov, 2025a]

$$|z>_{KP} = \hat{D}(z)\,|\,0> \equiv \frac{1}{\sqrt{{}_pN_q(|\,z\,|^2)}} \#\, {}_pF_q(z\hat{\mathcal{A}}_+)\; {}_pF_q(-z^*\hat{\mathcal{A}}_-) \#|\,0> \tag{3.59}$$

Since $\hat{\mathcal{A}}_- \,|\, 0>=0$, we can drop the second exponential, so the definition will become

$$|z>_{KP} \equiv \frac{1}{\sqrt{{}_pN_q(|\,z\,|^2)}}\; {}_pF_q(z\hat{\mathcal{A}}_+)\,|\,0> \tag{3.60}$$

From the definition of the generalized hypergeometric function and using the result of the action of the creation operator on the vacuum state, we obtain

$$|z>_{KP} \equiv \frac{1}{\sqrt{{}_pN_q(|\,z\,|^2)}} \sum_{n=0}^{\infty} \frac{z^n}{{}_p\rho_q(n)} \left(\hat{\mathcal{A}}_+\right)^n |\,0> = \frac{1}{\sqrt{{}_pN_q(|\,z\,|^2)}} \sum_{n=0}^{\infty} \sqrt{{}_p\rho_q(n)}\; z^n\,|\,n> \tag{3.61}$$

If we introduce a new notations

$${}_q\rho_p(n) = \frac{(n!)^2}{{}_p\rho_q(n)} \tag{3.62}$$

$${}_qF_p(|\,z\,|^2) = \sum_{n=0}^{n_{\max}\leq\ \infty} \frac{1}{{}_q\rho_p(n)} (|\,z\,|^2)^n = {}_qF_p\left(\left\{\left(\frac{b_j}{B_j}\right)_n\right\}_1^q ; \left\{\left(\frac{a_i}{A_i}\right)_n\right\}_1^p ; \frac{B_j}{A_i}\,|\,z\,|^2\right) \tag{3.63}$$

and using the normalization condition ${}_{KP}< z\,|\,z>_{KP} = 1$, the final expression of these generalized coherent states becomes

$$|z>_{KP} = \frac{1}{\sqrt{{}_qF_p(|\,z\,|^2)}} \sum_{n=0}^{n_{\max}\leq\ \infty} \frac{z^n}{\sqrt{{}_q\rho_p(n)}}\,|\,n> \tag{3.64}$$

From a mathematical point of view, the expression of these generalized coherent states (called the “Klauder-Perelomov” kind) is identical to the one obtained according to the previous definition (called “Barut-Girardello” type), the only difference being the interchange of the indices $p \leftrightarrow q$ and the sets of numbers $\{a_i\}_1^p \leftrightarrow \{b_j\}_1^q$ . This is a consequence of the existence of a duality between the two types of generalized coherent states, which we examined in a previous paper [Popov, 2025b]. For this reason we will reproduce here, without demonstration, the main results concerning generalized coherent states of the Klauder-Perelomov type. In fact, since the mathematical expressions are essentially the same, the demonstrations are similar to those for generalized coherent states of the Barut-Girardello type. For the same reason, in order not to burden the writing of formulas, we will attach the index KP only for generalized coherent states of the Klauder-Perelomov type. So, using the DOOT technique, the results are the following:

The projector

$$|\,z>_{KP\;KP}< z\,| = \frac{1}{{}_qF_p(|\,z\,|^2)} \# \frac{{}_qF_p(z\hat{\mathcal{A}}_+)\,{}_qF_p(z^*\hat{\mathcal{A}}_-)}{{}_pF_q(\hat{\mathcal{A}}_+\hat{\mathcal{A}}_-)} \# \tag{3.65}$$

where we observe the fact that the vacuum projector of the generalized coherent states of the Barut-Girardello type appeared, which can be expressed also through the product of pair operators $\#\hat{\mathcal{A}}_+\hat{\mathcal{A}}_-\#$ according to the following:

$$\hat{I} = \sum_{n=0}^{n_{\max}\leq\ \infty} |\, n >< n\,| = \sum_{n=0}^{n_{\max}\leq\ \infty} \frac{1}{{}_p\rho_q(n)} \left(\hat{\mathcal{A}}_+\right)^n |\,0 >< 0\,| \left(\hat{\mathcal{A}}_-\right)^n = |\,0 >< 0\,| \sum_{n=0}^{n_{\max}\leq\ \infty} \frac{\#(\hat{\mathcal{A}}_+\hat{\mathcal{A}}_-)^n\#}{{}_p\rho_q(n)} = \tag{3.66}$$
$$= |\,0 >< 0\,|\#\,{}_pF_q(\hat{\mathcal{A}}_+\hat{\mathcal{A}}_-)\#$$

Consequently, the vacuum projector is identical to that of generalized coherent states of the Barut-Girardello type

$$|\,0 >< 0\,| = \frac{1}{\#\,{}_pF_q(\hat{\mathcal{A}}_+\hat{\mathcal{A}}_-)\#} \tag{3.67}$$

The integration measure is then

$$d_{KP}(z) = \Gamma(b/a)\frac{d\varphi}{2\pi} d(|\,z\,|^2)\,{}_qF_p(|\,z\,|^2)\, G_{q,\,p+1}^{\,p+1,0}\left( \frac{\prod_{j=1}^{q} B_j}{\prod_{i=1}^{p} A_i} |\,z\,|^2 \,\middle|\, \begin{matrix} / \,; & \left\{\frac{b_j}{B_j} - 1\right\}_1^q \\ 0,\left\{\frac{a_i}{A_i} - 1\right\}_1^p ; & / \end{matrix} \right) \tag{3.68}$$

It is interesting to note another aspect of the duality between generalized coherent states of the Barut-Girardello and Klauder-Perelomov type: the latter also satisfy the same definition as the former, $\hat{\mathcal{A}}_-\,|\,z>_{KP} = z\,|\,z>_{KP}$ .

And another important observation: at the harmonic limit the hypergeometric function will become an exponential, so that we will obtain, as a particular case, the definition used by Klauder and Perelomov.

$$\lim_{\substack{p=q \\ \{a_i\}=\{b_j\}}} {}_qF_p(x) = \lim_{f(\hat{\mathcal{N}})\to 1} \sum_{n=0}^{\infty} \frac{1}{{}_q\rho_p(n)} x^n = \sum_{n=0}^{\infty} \frac{x^n}{n!} = e^x \tag{3.69}$$

Then, using the DOOT rules, we have

$$|z>_{KP} = \lim_{\substack{p=q \\ \{a_i\}=\{b_j\}}} \hat{\mathsf{D}}(z)\,|\,0> \equiv \lim_{\substack{p=q \\ \{a_i\}=\{b_j\}}} \frac{1}{\sqrt{{}_qN_p(|\,z\,|^2)}} \#\,{}_pF_q(z\,\hat{\mathcal{A}}_+)\ {}_pF_q(-z^*\,\hat{\mathcal{A}}_-)\#|\,0> =$$
$$= \frac{1}{\sqrt{{}_qN_p(|\,z\,|^2)}} \#e^{z\hat{\mathcal{A}}_+} e^{-z^*\hat{\mathcal{A}}_-}\#|\,0> = \frac{1}{\sqrt{{}_qN_p(|\,z\,|^2)}} e^{z\hat{\mathcal{A}}_+}\,|\,0> \tag{3.70}$$

Expanding the exponential in the last line in a power series and taking into account the result of the action of the creation operator on the vacuum state, we will obtain:

$$|z>_{KP} = \frac{1}{\sqrt{{}_qN_p(|\,z\,|^2)}} e^{z\hat{\mathcal{A}}_+}\,|\,0> = \frac{1}{\sqrt{{}_qN_p(|\,z\,|^2)}} \sum_{n=0}^{\infty} \frac{z^n}{n!} \left(\hat{\mathcal{A}}_+\right)^n |\,0> =$$
$$= \frac{1}{\sqrt{{}_qN_p(|\,z\,|^2)}} \sum_{n=0}^{\infty} \frac{z^n}{n!} \sqrt{{}_p\rho_q(n)}\,|\,n> \equiv \frac{1}{\sqrt{{}_qN_p(|\,z\,|^2)}} \sum_{n=0}^{\infty} \frac{z^n}{\sqrt{{}_q\rho_p(n)}} |\,n> \tag{3.71}$$

from which, using the normalization condition, the correct expression for the normalization function results ${}_qN_p(|z|^2)={}_qF_p(|z|^2)$.

***Definition III:*** **Generalized coherent states related directly to the system's Hamiltonian.** These states were introduced in 1999 by J. P. Gazeau and J. R. Klauder [Gazeau, 1999] and hence the name Gazeau-Klauder coherent states (GK-CSs). These are nonspreading and temporally stable states, directly related to the Hamiltonian and labeled by action-angle variables $0 \le J \le \infty$ and $-\infty \le \gamma \le +\infty$. For a Hamiltonian $\hat{\mathcal{H}}$ with dimensionless eigenvalues ordered as $0 = e(0) < e(1) < e(2) < ... < e(n)$, the GK-CSs are defined independently, as

$$|J,\gamma> = \frac{1}{\sqrt{{}_pF_q(J)}}\sum_{n=0}^{\infty}\frac{\left(\sqrt{J}\right)^n e^{-\mathrm{i}e(n)\gamma}}{\sqrt{{}_p\rho_q(n)}}|n> \tag{3.72}$$

These coherent states are generalized and can be obtained if the following steps are taken, in accordance with the rules of the DOOT technique [Popov, 2016]:

1.) It is assumed that the expression of the energy spectrum $E_n = \hbar\omega\, e(n)$ is known, as well as the action of the pair of operators $\hat{\mathcal{A}}_+$ and $\hat{\mathcal{A}}_-$ on the Fock basis vectors.

2.) We use the following definition, like Barut-Girardello coherent states, but for the set of coherent states with real variable $J$, where $|J> \equiv |J, \gamma = 0>$:

$$\hat{\mathcal{A}}_-|J> = \sqrt{J}\,|J> \tag{3.73}$$

3.) We expand the states $|J>$ into the Fock-vectors basis $|J> = \sum_{n=0}^{\infty} c_n(J)|n>$

4.) Through a recurrence relation, the connection between the coefficients is reached

$$c_n(J) = c_0(J)\frac{\left(\sqrt{J}\right)^n}{\sqrt{{}_p\rho_q(n)}} \tag{3.74}$$

5.) The normalization function is obtained as usually, from $<J|J>=1$

$$N_{GK}(J) = \sum_{n=0}^{\infty}\frac{J^n}{{}_p\rho_q(n)} = {}_pF_q(J) \tag{3.75}$$

6.) We act on the state $|J>$ with the exponential operator $\exp\left(-\mathrm{i}\,\hat{\mathcal{H}}\,t\right)$

$$|J,\gamma> \equiv \exp\left(-\mathrm{i}\hat{\mathcal{H}}\,t\right)|J> = \frac{1}{\sqrt{{}_pF_q(J)}}\sum_{n=0}^{\infty}\frac{\left(\sqrt{J}\right)^n}{\sqrt{{}_p\rho_q(n)}}\exp\left(-\mathrm{i}\hat{\mathcal{H}}\,t\right)|n> \tag{3.76}$$

7.) Finally, we obtain the searched expression that is the GK-CSs

$$|J,\gamma> = \frac{1}{\sqrt{{}_pF_q(J)}}\sum_{n=0}^{\infty}\frac{\left(\sqrt{J}\right)^n e^{-\mathrm{i}e(n)\gamma}}{\sqrt{{}_p\rho_q(n)}}|n> \tag{3.77}$$

In other words, the GK-CSs introduced by Gazeau and Klauder [Gazeau, 1999] can be obtained by applying the exponential operator $\#\exp\left(-\mathrm{i}\hat{\mathcal{H}}\,t\right)\# \equiv \#\exp\left(-\mathrm{i}\hat{\mathcal{A}}_+\hat{\mathcal{A}}_-\,t\right)\#$ on the BG-CSs depending only on the real positive variable $|J>$ and considering the rules of DOOT.

## 4. Mixed (thermal) states

Let us now consider a quantum system (which we call "quantum gas of oscillators") that is in thermodynamic equilibrium with the environment (called "bath"). If the gas only exchanges energy with the environment, then it obeys the canonical distribution. Its states will be mixed states characterized by the normalized canonical density operator and the partition function.

$$\rho = \frac{1}{Z(\beta)} e^{-\beta \hat{\mathcal{H}}} = \frac{1}{Z(\beta)} \sum_{n=0}^{\infty} e^{-\beta \hbar \omega e(n)} \,|\, n >< n \,| \quad , \quad Z(\beta) = \sum_{n=0}^{\infty} e^{-\beta \hbar \omega e(n)} \tag{4.1}$$

Using the result of the creation operator action on the vacuum state

$$\left(\hat{\mathcal{A}}_{+}\right)^{n} |\, 0 > = \sqrt{{}_{p}\rho_{q}(n)}\, |\, n > \tag{4.2}$$

as well as the DOOT expression for the vacuum operator, the density operator expression becomes

$$\rho = \frac{1}{Z(\beta)} \# \frac{1}{{}_{p}F_{q}\left(\hat{\mathcal{A}}_{+}\hat{\mathcal{A}}_{-}\right)} \sum_{n=0}^{\infty} e^{-\beta \hbar \omega \hat{\mathcal{A}}_{+}\hat{\mathcal{A}}_{-}} \frac{\#\left(\hat{\mathcal{A}}_{+}\hat{\mathcal{A}}_{-}\right)^{n}}{{}_{p}\rho_{q}(n)} \# \tag{4.3}$$

On the other hand, the expression of Husimi's distribution function [Husimi, 1940] is defined as being equal to the diagonal elements of the density operator in the representation of coherent states

$$Q\left(|\, z\,|^{2}\right) \equiv < \rho >_{z} = \frac{1}{Z(\beta)} \frac{1}{{}_{p}F_{q}\left(|\, z\,|^{2}\right)} \sum_{n=0}^{\infty} e^{-\beta \hbar \omega e(n)} \frac{\left(|\, z\,|^{2}\right)^{n}}{{}_{p}\rho_{q}(n)} \tag{4.4}$$

On the other hand, any density operator admits a spectral decomposition, i.e. a decomposition with respect to the coherent states projectors:

$$\rho = \int d\mu(z) P\left(|\, z\,|^{2}\right) |\, z >< z\,| =$$

$$= \Gamma\left(\frac{a}{A} \,/\, \frac{b}{B}\right) \int_{0}^{2\pi} \frac{d\varphi}{2\pi} \int_{0}^{R_c \le \infty} d\left(\frac{\prod_{i=1}^{p} A_i}{\prod_{j=1}^{q} B_j} |\, z\,|^{2}\right) G_{p,q+1}^{q+1,0}\left(\frac{\prod_{i=1}^{p} A_i}{\prod_{j=1}^{q} B_j} |\, z\,|^{2} \middle| \begin{array}{ll} / ; & \left\{\frac{a_i}{A_i} - 1\right\}_{1}^{p} \\ 0, \left\{\frac{b_j}{B_j} - 1\right\}_{1}^{q} ; & / \end{array}\right) \times \tag{4.5}$$

$$\times\, {}_{p}F_{q}\left(|\, z\,|^{2}\right) P\left(|\, z\,|^{2}\right) |\, z >< z\,|$$

where $P\left(|\, z\,|^{2}\right)$ is a quasi-distribution (in the sense that it can also take negative values on certain ranges of the $|\, z\,|^{2}$ variable).

Using the expression for the projector expansion of coherent states $|\, z >< z\,|$ in terms of Fock vectors $|\, n >< n\,|$ and taking into account that the angular integral has the value $\left(|\, z\,|^{2}\right)^{n} \delta_{nm}$, we will obtain that the spectral development of the density operator reduces to

$$\rho = \Gamma\left(\frac{a}{A} / \frac{b}{B}\right) \sum_{n=0}^{\infty} | n >< n | \frac{1}{{}_p\rho_q(n)} \times$$

$$\times \int_0^{R_c \le \infty} d\left(\frac{\prod_{i=1}^{p} A_i}{\prod_{j=1}^{q} B_j} | z |^2\right) G_{p,q+1}^{q+1,0}\left(\frac{\prod_{i=1}^{p} A_i}{\prod_{j=1}^{q} B_j} | z |^2 \middle| \begin{matrix} / ; & \left\{\frac{a_i}{A_i} - 1\right\}_1^p \\ 0, \left\{\frac{b_j}{B_j} - 1\right\}_1^q ; & / \end{matrix}\right) P\left(| z |^2\right)\left(| z |^2\right)^n \tag{4.6}$$

The condition for having equality between the two expressions of the density operator, i.e. Eqs. (4.1) and (4.6) is

$$\int_0^{R_c \le \infty} d\left(\frac{\prod_{i=1}^{p} A_i}{\prod_{j=1}^{q} B_j} | z |^2\right) G_{p,q+1}^{q+1,0}\left(\frac{\prod_{i=1}^{p} A_i}{\prod_{j=1}^{q} B_j} | z |^2 \middle| \begin{matrix} / ; & \left\{\frac{a_i}{A_i} - 1\right\}_1^p \\ 0, \left\{\frac{b_j}{B_j} - 1\right\}_1^q ; & / \end{matrix}\right) P\left(| z |^2\right)\left(| z |^2\right)^n = \tag{4.7}$$

$$= \frac{1}{Z(\beta)} \frac{1}{\Gamma\left(\frac{a}{A} / \frac{b}{B}\right)} e^{-\beta \hbar \omega e(n)} \; {}_p\rho_q(n)$$

If we introduce the unknown function as

$$\tilde{P}\left(| z |^2\right) \equiv G_{p,q+1}^{q+1,0}(\ldots) \; P\left(| z |^2\right) \tag{4.8}$$

simultaneously with index change $n = s - 1$ , we will arrive at an integral equation:

$$\int_0^{R_c \le \infty} d\left(| z |^2\right) \tilde{P}\left(| z |^2\right)\left(| z |^2\right)^{s-1} = \frac{1}{Z(\beta)} \frac{1}{\Gamma\left(\frac{a}{A} / \frac{b}{B}\right)} e^{-\beta \hbar \omega e(s-1)} \; {}_p\rho_q(s-1) \tag{4.9}$$

From a mathematical point of view, such an equation is called the moment problem. As regarded the coherent states, two types of moments arise: the Stieltjes moment problem, for the entire positive axis $(0, +\infty)$, and the Hausdorff moment problem for a bounded interval, $(0, 1)$ [Schmüdgen, 2017] .

$$R_c = \lim_{n \to \infty} \frac{{}_q\rho_p(n)}{{}_q\rho_p(n+1)} = \begin{cases} \infty & \text{Stieltjes momemt problem} \\ < \infty & \text{Haussdorff moment problem} \end{cases} \tag{4.10}$$

From a practical point of view, the diagonal representation of the density operator is useful in calculating the expected mean thermal value of an operator $\hat{O}$ that characterizes the quantum system studied.

$$< \hat{O} > = \mathrm{Tr}\left(\rho \hat{O}\right) = \int d\,\mu(\sigma) < \sigma | \rho \hat{O} | \sigma > = \int d\mu(z) P\left(| z |^2\right) < z | \hat{O} | z > \tag{4.11}$$

Substituting the expressions for the integration measure and the respective coherent states, we will obtain the final result, i.e. the expression of the thermal expectation value in the representation of Fock vectors.

$$< \hat{O} > = \frac{1}{Z(\beta)} \sum_{n=0}^{\infty} e^{-\beta \hbar \omega e(n)} < n | \hat{O} | n > \tag{4.12}$$

From this, two conclusions can be drawn:

1.) The calculation of the thermal expected value does not depend on the representation used, so the representation that involves simpler calculations will be chosen.

2.) In the expression of the thermal expected value, regardless of the nature of operators, only the diagonal elements of the operators intervene, therefore they contribute.

For the particular case $\hat{O} = \#\left(\hat{\mathcal{A}}_+ \hat{\mathcal{A}}_-\right)^m \#$, we will have

$$< \#\left(\hat{\mathcal{A}}_+ \hat{\mathcal{A}}_-\right)^m \# > = \frac{1}{Z(\beta)} \sum_{n=0}^{\infty} e^{-\beta\hbar\omega e(n)} < n | \#\left(\hat{\mathcal{A}}_+ \hat{\mathcal{A}}_-\right)^m \# | n > = \frac{1}{Z(\beta)} \sum_{n=0}^{\infty} e^{-\beta\hbar\omega e(n)} \left[e(n)\right]^m \tag{4.13}$$

For simplicity, let us introduce the notation $X \equiv e^{-\beta\hbar\omega}$, and $\vartheta_X \equiv X \dfrac{\partial}{\partial X}$ so we will have

$$< \#\left(\hat{\mathcal{A}}_+ \hat{\mathcal{A}}_-\right)^m \# > = \frac{1}{Z(\beta)} \sum_{n=0}^{\infty} X^{e(n)} \left[e(n)\right]^m = \frac{1}{Z(\beta)} \left(\vartheta_X\right)^m Z(\beta) \tag{4.14}$$

Now we can express the generalized counterpart of the Mandel parameter, which we introduced in a previous paper [Popov, 2003]. Let's remember that the Mandel Q parameter measures the departure of the occupation number distribution from Poissonian statistics.

$$Q_{th}(\beta) = \frac{< \#\left(\hat{\mathcal{A}}_+ \hat{\mathcal{A}}_-\right)^2 \# > - \left(< \#\left(\hat{\mathcal{A}}_+ \hat{\mathcal{A}}_-\right) \# >\right)^2}{< \#\left(\hat{\mathcal{A}}_+ \hat{\mathcal{A}}_-\right) \# >} - 1 = \frac{\mathrm{Var}\left(< \#\left(\hat{\mathcal{A}}_+ \hat{\mathcal{A}}_-\right) \# >\right)}{\left(< \#\left(\hat{\mathcal{A}}_+ \hat{\mathcal{A}}_-\right) \# >\right)} - 1 \tag{4.15}$$

The variance of an operator $\hat{\mathcal{X}}$, is a positive quantity, or zero, i.e.

$$\mathrm{Var}\left(\hat{\mathcal{X}}\right) \equiv < \left(\hat{\mathcal{X}} - < \hat{\mathcal{X}} >\right)^2 > = < \hat{\mathcal{X}}^2 > - \left(< \hat{\mathcal{X}} >\right)^2 \geq 0 \tag{4.16}$$

so we can evaluate the sign of Mandel's parameter:

$$Q_{th}(\beta) = \frac{\frac{1}{Z(\beta)} \left(\vartheta_X\right)^2 Z(\beta) - \left[\frac{1}{Z(\beta)} \vartheta_X Z(\beta)\right]^2}{\frac{1}{Z(\beta)} \vartheta_X Z(\beta)} - 1 = \frac{\sum_{n=0}^{\infty} X^{e(n)} \left[e(n)\right]^2}{\sum_{n=0}^{\infty} X^{e(n)} e(n)} - \frac{\left[\sum_{n=0}^{\infty} X^{e(n)} e(n)\right]^2}{\sum_{n=0}^{\infty} X^{e(n)}} - 1 \tag{4.17}$$

Because $e(n) = n\, f(n)$, we have $\sum_{n=0}^{\infty} X^{e(n)} \left[e(n)\right]^2 > \sum_{n=0}^{\infty} X^{e(n)} e(n)$, and we obtain

$$Q_{th}(\beta) < 1 - \frac{\sum_{n=0}^{\infty} X^{e(n)} e(n)}{\sum_{n=0}^{\infty} X^{e(n)}} - 1 < 0 \tag{4.18}$$

i.e. the non-linear generalized thermal states display nonclassical (sub-Poissonian) distribution function [Dhanalaksmi, 2013],[Pennini, 2014].

## 5. Coherent states for systems with continuous energy spectrum

Let us now turn our attention to quantum systems that have a mixed, both discrete and continuous energy spectrum (such as, for example, the Morse oscillator). The continuous spectrum arises in the situation if the variable $x$ of the potential $V(x)$ becomes very large, $x \to \infty$. These systems possess a discrete part of the spectrum, consisting of a finite number of bound states, as well as a continuous part, consisting of an infinity of unbound (or diffusion) states. For this type of systems, the closure relation will be written [Messiah, 1967]

$$\sum_{n=0}^{n_{\max}} |n><n| + \int_0^{\infty} dE |E><E| = 1 \quad . \tag{5.1}$$

where $|E>$ are eigenstates of the dimensionless Hamiltonian $\hat{\mathcal{H}}$ with a non-degenerate continuous spectrum, $\hat{\mathcal{H}}|E>=E|E>$. These states are formal delta-function normalized, i.e. $<E|E'>=\delta(E-E')$. The closure relation for continuous part of spectrum is then

$$\int_0^{\infty} dE |E><E| = 1_c \quad . \tag{5.2}$$

We introduce the following limit (for brevity called *the discrete – continuous limit* $d \to c$), defined as (see, also, [Popov, 2016a]) which makes the connection between the characteristic features of the continuous spectrum $X_c(E)$ and those corresponding to the discrete spectrum $X_d(n)$.

$$X_c(E) = \lim_{\substack{n\to E \\ n_{\max}\to\infty \\ z\to Z \\ p=q \\ \{a_i\}=\{b_j\} \\ \{A_i\}=\{B_j\}=1 \\ \sum\to\int}} X_d(n) \equiv \lim_{d\to c} X_d(n) \quad , \qquad \sum_{n=0}^{n_{\max}} X_d(n) \to \int_0^{\infty} dE\, X_c(E)(E) \tag{5.3}$$

To distinguish between the entities of the discrete spectrum $X_d(n)$ and those of the continuous spectrum $X_c(E)$, instead of the variable $z = |z| \exp(\,\mathrm{i}\varphi)$ we will use the variable $Z \equiv |Z| e^{-\mathrm{i}\gamma}$ with $0 \le |Z| \le R_c \le \infty$, $-\infty \le \gamma \le +\infty$

The conditions in the definition of the limit above, i.e. the equality of the indices $p = q$ and the sets of real numbers $\left\{ \frac{a_i}{A_i} \right\}_1^p = \left\{ \frac{b_j}{B_j} \right\}_1^q$, with the substitution $n \to E$, lead to

$$\lim_{d\to c} \rho_{p,q}(n) = \lim_{d\to c} n! \frac{\prod_{j=1}^{q} \left( \frac{b_j}{B_j} \right)_n}{\prod_{i=1}^{p} \left( \frac{a_i}{A_i} \right)_n} = \Gamma(E+1) \tag{5.4}$$

$$\lim_{d\to c} {}_pF_q\left( \left\{ \frac{a_i}{A_i} \right\}_1^p ; \left\{ \frac{b_j}{B_j} \right\}_1^q ; \frac{A_i}{B_j} |z|^2 \right) = \lim_{d\to c} \sum_{n=0}^{\infty} \frac{1}{\rho_{p,q}(n)} (|z|^2)^n = \int_0^{\infty} dE \frac{(|Z|^2)^E}{\Gamma(E+1)} = \nu(|Z|^2) \tag{5.5}$$

Consequently, for the continuous spectrum case, the generalized normalization function ${}_pF_q(...)$ is replaced by the so-called $\nu$-function [Gradshteyn, 2007], defined as

$$\int_0^{\infty} dE \frac{(|Z|^2)^E}{\Gamma(E+1)} = \nu(|Z|^2) \tag{5.6}$$

Therefore, the expansion of the coherent states for the continuous spectrum in terms of energy eigenvectors are, see, [Gazeau, 1999], and [Inomata, 2005] (unprinted)

$$|Z> = \frac{1}{\sqrt{\nu\left(|Z|^2\right)}} \int_0^\infty dE \frac{Z^E}{\sqrt{\Gamma(E+1)}} |E> \quad . \tag{5.7}$$

Applying the discrete-continuous limit, we convince ourselves that we will have

$$\lim_{d \to c} |z> = |Z> \tag{5.8}$$

The structure of the coherent states for the continuous spectrum is the same for *all* quantum systems, i.e. they are independent of the potential which is involved in generating the discrete spectrum. It is not difficult to verify that these coherent states are continuous in label $Z$ and saturate the decomposition relation of the identity operator.

$$\int d\mu_c(Z) |Z><Z| = 1_c \tag{5.9}$$

The corresponding integration measure is

$$d\mu_c(Z) = \frac{d\gamma}{2\pi} d\left(|Z|^2\right) e^{-|Z|^2} \nu\left(|Z|^2\right) \tag{5.10}$$

Also, the coherent states of the continuous spectrum are temporally stable.

$$e^{-\mathrm{i}\omega t H} \; |Z> = |Z(t)> \qquad , \qquad Z(t) \equiv |Z| e^{-\mathrm{i}(\gamma + \omega t)} \text{ and } \omega = \text{const} . \tag{5.11}$$

The expectation value of an operator $\hat{O}$ in the $|Z>$ representation is

$$<\hat{O}>_Z \equiv <Z|\hat{O}|Z> = \frac{1}{\nu\left(|Z|^2\right)} \int_0^\infty dE\, dE' \frac{\left(Z^*\right)^E Z^{E'}}{\sqrt{\Gamma(E+1)\Gamma(E'+1)}} <E|\hat{O}|E'> \tag{5.12}$$

For the particular case $\hat{O} = \hat{\mathcal{H}}$, we obtain

$$<Z|\hat{\mathcal{H}}|Z> = \frac{1}{\nu\left(|Z|^2\right)} \int_0^\infty dE\, E \frac{\left(|Z|^2\right)^E}{\Gamma(E+1)} = |Z|^2 \frac{d}{d\left(|Z|^2\right)} \log \nu\left(|Z|^2\right) = |Z|^2 \equiv H\left(|Z|^2\right) \tag{5.13}$$

This is obtained using an interesting property of the $\nu$-function:

$$\frac{d}{d|Z|^2} \nu\left(|Z|^2\right) = \int_0^\infty dE\, E \frac{\left(|Z|^2\right)^{E-1}}{\Gamma(E+1)} = \int_0^\infty d(E-1) \frac{\left(|Z|^2\right)^{E-1}}{\Gamma[(E-1)+1]} = \nu\left(|Z|^2\right) \tag{5.14}$$

as well as the property $\Gamma(E+1) = E\,\Gamma(E)$ (remember, the $\nu$-function is defined *only* on the interval $(0\,,\,+\infty)$). If we set $H\left(|Z|^2\right) = \omega J$, we obtain $|Z|^2 = \omega J = f(J)$. This is the fourth Klauder's requirement for coherent states, the so called action identity (see, [Gazeau, 1999]), and it is accomplished also for coherent states of continuous spectrum.

By concluding, the coherent states for continuous spectrum $|Z>$ satisfy all Klauder's minimal requirements [Gazeau, 1999]: (a) continuity in the complex variable; (b) resolution of unity; (c) temporal stability and (d) action identity.

By analogy with the pair of creation and annihilation operators for the discrete spectrum,

$$\hat{\mathcal{A}}_- = \sum_{n=0}^\infty \sqrt{e(n+1)}\, |n><n+1| , \qquad \hat{\mathcal{A}}_+ = \sum_{n=0}^\infty \sqrt{e(n+1)}\, |n+1><n| , \tag{5.15}$$

we introduce a similar pair of adjoint operators for the continuous spectrum, $\hat{\mathcal{A}}_-^{(c)}$ and $\hat{\mathcal{A}}_+^{(c)}$, which, after applying the *discrete – continuous limit* $d \to c$**,** satisfies the equations

$$\hat{\mathcal{A}}_{-}^{(c)} = \lim_{d\to c}\hat{\mathcal{A}}_{-} = \int_0^\infty dE\,\sqrt{E+1}\,|\,E><E+1\,|\ ,\qquad \hat{\mathcal{A}}_{+}^{(c)} = \lim_{d\to c}\hat{\mathcal{A}}_{+} = \int_0^\infty dE\,\sqrt{E+1}\,|\,E+1><E\,| \tag{5.16}$$

$$\lim_{d\to c} {}_p e_q(n) = \lim_{d\to c} n\,\frac{\prod_{j=1}^{q}\left[b_j + B_j(n-1)\right]}{\prod_{i=1}^{p}\left[a_i + A_i(n-1)\right]} = E \tag{5.17}$$

We obtain successively how these operators act on the eigenvectors $|\,E>$

$$\hat{\mathcal{A}}_{-}^{(c)}\,|\,E> = \int_0^\infty dE'\sqrt{E'+1}\,|\,E'><E'+1\,|\,E> = \int_0^\infty dE'\sqrt{E'+1}\,|\,E'>\delta_{E'+1,E} = \sqrt{E}\,|\,E-1> \tag{5.18}$$

$$\hat{\mathcal{A}}_{+}^{(c)}\,|\,E> = \int_0^\infty dE'\sqrt{E'+1}\,|\,E'+1><E'\,|\,E> = \int_0^\infty dE'\sqrt{E'+1}\,|\,E'+1>\delta_{E',E} = \sqrt{E+1}\,|\,E+1> \tag{5.19}$$

$$<E\,|\left(\hat{\mathcal{A}}_{+}^{(c)}\hat{\mathcal{A}}_{-}^{(c)}\right)^m|\,E> = E^m\quad ,\qquad \left(\hat{\mathcal{A}}_{+}^{(c)}\hat{\mathcal{A}}_{-}^{(c)}\right)^m = \int_0^\infty dE\,E^m\,|\,E><E\,| \tag{5.20}$$

We also find the following useful equality, obtained using the discrete – continuous limit $d\to c$:

$$\lim_{d\to c}\left(\hat{\mathcal{A}}_{+}\right)^n|\,0> = \lim_{d\to c}\prod_{j=0}^{n-1} f_{p,q}(j)\,|\,n> = \lim_{d\to c}\sqrt{\rho_{p,q}(n)}\,|\,n> = \sqrt{\Gamma(E+1)}\,|\,E> = \left(\hat{\mathcal{A}}_{+}^{(c)}\right)^E|\,0> \tag{5.21}$$

as well as their adjoint conterpart

$$|\,E> = \frac{1}{\sqrt{\Gamma(E+1)}}\left(\hat{\mathcal{A}}_{+}^{(c)}\right)^E|\,0>\quad ,\qquad <E\,| = \frac{1}{\sqrt{\Gamma(E+1)}}<0\,|\left(\hat{\mathcal{A}}_{-}^{(c)}\right)^E \tag{5.22}$$

The expectation values in the $|\,Z>$ representation, for an operator $\hat{O}$ is

$$<\hat{O}>_Z = \frac{1}{\nu\left(|\,Z\,|^2\right)}\int_0^\infty dE\,dE'\,\frac{\left(Z^*\right)^E Z^{E'}}{\sqrt{\Gamma(E+1)\Gamma(E'+1)}}<E\,|\,\hat{O}\,|\,E'> \tag{5.23}$$

If $\hat{O} = \hat{\mathcal{A}}_{+}^{(c)}\hat{\mathcal{A}}_{-}^{(c)}$ we obtain the same result as for discrete spectrum

$$<\left(\hat{\mathcal{A}}_{+}^{(c)}\hat{\mathcal{A}}_{-}^{(c)}\right)^m>_Z = \frac{1}{\nu\left(|\,Z\,|^2\right)}\left(|\,Z\,|^2\,\frac{d}{d\,|\,Z\,|^2}\right)^m\nu\left(|\,Z\,|^2\right) = \left(|\,Z\,|^2\right)^m \tag{5.24}$$

i.e the same value as for the Hamiltonian $<Z\,|\,\hat{\mathcal{H}}\,|\,Z>$. This means the following equality: $\hat{\mathcal{H}} = \hat{\mathcal{A}}_{+}^{(c)}\hat{\mathcal{A}}_{-}^{(c)}$.

Consequently, the value of the Mandel parameter [Mandel, 1994] in the $|\,Z>$ representation shows that the coherent states for continuous spectrum obey the sub-Poissonian distribution.

$$Q_Z^{(c)} = \frac{<\left(\hat{\mathcal{A}}_{+}^{(c)}\hat{\mathcal{A}}_{-}^{(c)}\right)^2>_Z - \left(<\hat{\mathcal{A}}_{+}^{(c)}\hat{\mathcal{A}}_{-}^{(c)}>_Z\right)^2}{<\hat{\mathcal{A}}_{+}^{(c)}\hat{\mathcal{A}}_{-}^{(c)}>_Z} - 1 = -1 \tag{5.25}$$

Applying the discrete – continuous limit leads us to the expression of the particle number operator for the continuous spectrum $\hat{\mathcal{N}}^{(c)}$

$$\hat{\mathcal{N}}^{(c)} = \lim_{d\to c}\hat{\mathcal{N}} = \lim_{d\to c}\sum_{n=0}^{\infty} n | n >< n | = \int_0^{\infty} dE\, E\, | E >< E | = \hat{\mathcal{A}}_+^{(c)}\hat{\mathcal{A}}_-^{(c)} = \hat{\mathcal{H}} \tag{5.26}$$

The expectation value of an integer power $s$ of number operator $\hat{\mathcal{N}}^{(c)}$ is

$$< E | \left(\hat{\mathcal{N}}^{(c)}\right)^s | E > \equiv < \left(\hat{\mathcal{N}}^{(c)}\right)^s >_E = < E | \left(\hat{\mathcal{A}}_+^{(c)}\hat{\mathcal{A}}_-^{(c)}\right)^s | E > = E^s \tag{5.27}$$

Consequently, the value of the Mandel parameter in the $| E >$ representation shows that the coherent states for continuous spectrum obey the sub-Poissonian distribution.

$$Q_E^{(c)} = \frac{< \left(\hat{\mathcal{N}}^{(c)}\right)^2 >_E - \left(< \hat{\mathcal{N}}^{(c)} >_E\right)^2}{< \hat{\mathcal{N}}^{(c)} >_E} - 1 = -1 \tag{5.28}$$

Using the discrete- continuous limit, we can adapt the DOOT rules, formulated in [Popov, 2015b] also for the case of continuous spectrum. Because the operators $\hat{\mathcal{A}}_-^{(c)}$ and $\hat{\mathcal{A}}_+^{(c)}$ act as *lowering* and *raising* operators, their product operator in the normal ordered manner $\hat{\mathcal{A}}_+^{(c)}\hat{\mathcal{A}}_-^{(c)}$ is a diagonal operator in the energy eigenvectors basis $| E >$. Specifically, the projector $| 0 >< 0 |$ of the normalized vacuum state $| 0 >$, in the frame of DOOT, is

$$| 0 >< 0 | = \lim_{d\to c} \# \frac{1}{{}_pF_q\left(\left\{\left(\frac{a_i}{A_i}\right)_n\right\}_1^p ; \left\{\left(\frac{b_j}{B_j}\right)_n\right\}_1^q ; \frac{A_i}{B_j}\hat{\mathcal{A}}_+\hat{\mathcal{A}}_-\right)} \# = \# \frac{1}{\nu\left(\hat{\mathcal{A}}_+^{(c)}\hat{\mathcal{A}}_-^{(c)}\right)} \# \tag{5.29}$$

The following equalities are also valid:

$$\# \mathcal{F}(\hat{\mathcal{A}}_+^{(c)}\hat{\mathcal{A}}_-^{(c)}) \# | E > = \mathcal{F}(E) | E > \;,\quad \# \mathcal{F}(\hat{\mathcal{A}}_+^{(c)}\hat{\mathcal{A}}_-^{(c)}) \# = \int_0^{\infty} dE\, \mathcal{F}(E)\, | E >< E | \tag{5.30}$$

Using the above relations and the DOOT rules, the coherent states for the continuous spectrum can be written in the following operatorial manner:

$$| Z > = \frac{1}{\sqrt{\nu\left(| Z |^2\right)}}\nu\left(Z\hat{\mathcal{A}}_+^{(c)}\right) | 0 > \;,\quad < Z | = \frac{1}{\sqrt{\nu\left(| Z |^2\right)}} < 0 | \nu\left(Z^*\hat{\mathcal{A}}_-^{(c)}\right) \tag{5.31}$$

Negative value of the Mandel parameter show that the CSs $| Z >$ have the Poissonian behavior or statistics. The probability density function of the transition $| E >$ to state $| Z >$ is

$$\mathcal{P}_E\left(| Z |^2\right) = | < Z | E > |^2 = \frac{1}{\nu\left(| Z |^2\right)}\frac{\left(| Z |^2\right)^E}{\Gamma(E+1)} \tag{5.32}$$

If we apply the *inverse* limit, i.e. the continuous – discrete limit $c \to d$ we obtain just the Poisson probability density function for the discrete case:

$$\lim_{c\to d}\mathcal{P}_E\left(| Z |^2\right) \equiv \lim_{\substack{E\to n \\ \int\to\sum}} \mathcal{P}_E\left(| Z |^2\right) = \frac{1}{\sum_{n=0}^{\infty}\frac{\left(| Z |^2\right)^n}{n!}}\frac{\left(| Z |^2\right)^n}{n!} = e^{-|Z|^2}\frac{\left(| Z |^2\right)^n}{n!} \equiv \mathcal{P}_n\left(| Z |^2\right) \tag{5.33}$$

If we consider that the quantum system (of free particles, because the potential $V = 0$) is in the thermodynamic equilibrium with the environment at temperature $T = (k_B\,\beta)^{-1}$, then their states are mixed described by the normalized density operator obeying the canonical distribution

$$\rho_c = \lim_{d\to c}\rho = \frac{1}{Z_c(\beta)}\int_0^\infty dE\ e^{-\beta\hbar\omega E}\,|\,E><E\,| \tag{5.34}$$

The continuous partition function $Z_c(\beta)$ will be determined by normalizing density operator**:**

$$\mathrm{Tr}\rho_c = \int d\mu_c(Z')<Z'|\,\rho_c\,|Z'>=\frac{1}{Z_c(\beta)}\int_0^\infty dE\, e^{-\beta\hbar\omega E} = \frac{1}{Z_c(\beta)}\frac{1}{\beta\hbar\omega}=1 \tag{5.35}$$

and we obtain:

$$Z_c(\beta)=\frac{1}{\beta\hbar\omega}=\left(\log\frac{\bar{n}+1}{\bar{n}}\right)^{-1} \quad , \qquad \bar{n}=\left(e^{\beta\hbar\omega}-1\right)^{-1} \tag{5.36}$$

where $\bar{n}$ is the Bose-Einstein distribution function for oscillators with angular frequency $\omega$ .

Using the DOOT rules, the density operator becomes

$$\rho_c = \frac{1}{Z_c(\beta)}\#\frac{1}{\nu\left(\hat{\mathcal{A}}_+^{(c)}\hat{\mathcal{A}}_-^{(c)}\right)}\int_0^\infty dE\ \frac{1}{\Gamma(E+1)}\left(e^{-\beta\hbar\omega}\,\hat{\mathcal{A}}_+^{(c)}\hat{\mathcal{A}}_-^{(c)}\right)^E \#=\frac{1}{Z_c(\beta)}\#\frac{\nu\left(\frac{\bar{n}}{\bar{n}+1}\hat{\mathcal{A}}_+^{(c)}\hat{\mathcal{A}}_-^{(c)}\right)}{\nu\left(A_+A_-\right)}\# \tag{5.37}$$

The corresponding $Q$-distribution function is then

$$Q_c\left(|\,Z\,|^2\right)\equiv<\rho_c>_Z=\frac{1}{\nu\left(|\,Z\,|^2\right)}\int_0^\infty dE\,dE'\frac{\left(Z^*\right)^E Z^{E'}}{\sqrt{\Gamma\left(E+1\right)\Gamma\left(E'+1\right)}}<E|\,\rho_c\,|E'> \tag{5.38}$$

Because the elements of the density operator are

$$<E|\,\rho_c\,|E'>=\frac{1}{Z_c(\beta)}<E|\#e^{-\beta\hbar\omega\hat{\mathcal{A}}_+^{(c)}\hat{\mathcal{A}}_-^{(c)}}\#|E'>=\frac{1}{Z_c(\beta)}e^{-\beta\hbar\omega E}\delta(E-E') \tag{5.39}$$

finally, we obtain the normalized $Q$-distribution function

$$Q_c\left(|\,Z\,|^2\right)=\frac{1}{Z_c(\beta)}\frac{\nu\left(|\,Z\,|^2\ \frac{\bar{n}}{\bar{n}+1}\right)}{\nu\left(|\,Z\,|^2\right)} \quad , \qquad \int d\mu_c(Z)\,Q_c\left(|\,Z\,|^2\right)=1 \tag{5.40}$$

The diagonal expansion of the normalized canonical density operator $\rho_c$ in terms of the $|\,Z>$ - coherent states projectors is:

$$\rho_c = \int d\mu_c(Z)\,P_c(|\,Z\,|^2)|\,Z><Z\,| \tag{5.41}$$

and we must determine the $P$ - quasi-distribution function.

Using a standard procedure, as in the discrete case, we will obtain a correct expression for $P$ - quasi distribution function. However, this can be obtained more simply by using the succession of continuous – discrete limits $c\to d$

$$\widetilde{P}_c(|\,Z\,|^2)=\lim_{d\to c}\widetilde{P}\left(|\,z\,|^2\right)=G_{0,1}^{1,0}\left(|\,Z\,|^2\,|0\right)\ P_c\left(|\,Z\,|^2\right)=e^{-|Z|^2}P_c\left(|\,Z\,|^2\right) \tag{5.42}$$

$$\lim_{d\to c}\int_0^{R_c\le\infty}d\left(|\,z\,|^2\right)\ \widetilde{P}\left(|\,z\,|^2\right)\left(|\,z\,|^2\right)^{s-1}=\lim_{d\to c}\frac{1}{Z_c(\beta)}\frac{1}{\Gamma(a/b)}\ e^{-\beta\hbar\omega e(s-1)}\ {}_p\rho_q\left(s-1\right) \tag{5.43}$$

equality leading to

$$\lim_{d\to c}\int_0^{R_c\le\infty} d\left(|z|^2\right)\,\tilde{P}\left(|z|^2\right)\left(|z|^2\right)^{s-1} \equiv \int_0^{R_c\le\infty} d\left(|Z|^2\right)\,e^{-|Z|^2}P_c\left(|Z|^2\right)\left(|z|^2\right)^{s-1}$$

$$\lim_{d\to c}\frac{1}{\Gamma(a/b)}\,e^{-\beta\hbar\omega e(s-1)}\ {}_p\rho_q\,(s-1) = \frac{1}{Z(\beta)}\left(e^{-\beta\hbar\omega}\right)^{(s-1)}(s-1)! = \frac{1}{Z_c(\beta)}\left(e^{-\beta\hbar\omega}\right)^{(s-1)}\Gamma(s) \tag{5.44}$$

$$\int_0^{R_c\le\infty} d\left(|Z|^2\right)\,e^{-|Z|^2}P_c\left(|Z|^2\right)\left(|z|^2\right)^{s-1} = \frac{1}{Z_c(\beta)}\left(e^{-\beta\hbar\omega}\right)^{(s-1)}\Gamma(s) \tag{5.45}$$

In order to solving this moment problem we perform the function change $H_c\left(|Z|^2\right)\equiv e^{-|Z|^2}P_c\left(|Z|^2\right)$ .

$$\int_0^{R_c\le\infty} d\left(|Z|^2\right)\,H_c\left(|Z|^2\right)\left(|z|^2\right)^{s-1} = \frac{e^{\beta\hbar\omega}}{Z_c(\beta)}\frac{1}{\left(e^{\beta\hbar\omega}\right)^s}\Gamma(s) \tag{5.46}$$

$$H_c\left(|Z|^2\right) = \frac{e^{\beta\hbar\omega}}{Z(\beta)}G_{0,1}^{1,0}\left(e^{\beta\hbar\omega}\ |Z|^2\,\middle|\,0\right) = \frac{e^{\beta\hbar\omega}}{Z_c(\beta)}e^{-e^{\beta\hbar\omega}|Z|^2} \tag{5.47}$$

$$P_c\left(|Z|^2\right) = \beta\hbar\omega e^{\beta\hbar\omega}\ e^{-\left(e^{\beta\hbar\omega}-1\right)|Z|^2} = \beta\hbar\omega\,\frac{\bar{n}+1}{\bar{n}}e^{-\frac{1}{\bar{n}}|Z|^2} \tag{5.48}$$

i.e. the same expression as there of applying the inverse limit $d\to c$ to the quasi $P$-distribution.

It is not difficult to prove that the $P$ - quasi-distribution function is normalized to unity

$$\int d\mu_c(Z)P_c(|Z|^2) = 1 \tag{5.49}$$

Because the thermal expectation values (thermal averages) of the operators $\hat{O}$ which characterize the quantum system, does not depend on the representation, it can be expressed in the $|Z>$ - , as well as in the $|E>$ - representations, which finally leads to:

$$<\hat{O}> = \mathrm{Tr}(\#\rho_c\hat{O}\#) = \int_0^\infty dE <E|\#\rho_c\hat{O}\#|E> = \beta\hbar\omega\int_0^\infty dE\,e^{-\beta\hbar\omega E}\ <E|\#\hat{O}\#|E> \tag{5.50}$$

For integer $s$ - power of the number operator is

$$<\left(\hat{\mathcal{N}}^{(c)}\right)^s> = \frac{1}{Z_c(\beta)}\int_0^\infty dE\left(\beta\hbar\omega\right)^{-E} <E|\left(\hat{\mathcal{N}}^{(c)}\right)^s|E> = \beta\hbar\omega\int_0^\infty dE\,E^s\,e^{-\beta\hbar\omega E} = \frac{s!}{\left(\beta\hbar\omega\right)^s} \tag{5.51}$$

The thermal counterpart of the Mandel parameter is then a linear function with repect to the temperature $T$ :

$$Q_c(T) = \frac{<\left(\hat{\mathcal{N}}^{(c)}\right)^2>_E - \left(<\hat{\mathcal{N}}^{(c)}>_E\right)^2}{<\hat{\mathcal{N}}^{(c)}>_E} - 1 = \frac{1}{\beta\hbar\omega} - 1 = \frac{k_B}{\hbar\omega}T - 1 \tag{5.52}$$

and it can be connected with the Debye temperature $T_D$ :

$$Q_c(T) = \frac{T}{T_D} - 1 \quad , \quad T_D = \frac{\hbar\omega}{k_B} \tag{5.53}$$

The behavior of the thermal Mandel parameter $Q_c(T)$ as function of the equilibrium temperature $T$ can be summarized in the table below:

$$Q_c(T)\begin{cases} < T_D & \text{Sub - Poissonian statistics} \\ = T_D & \text{Poissonian statistics} \\ > T_D & \text{Super - Poissonian statistics} \end{cases}$$

## 6. Summary results for generalized coherent states of some quantum oscillators

### 6.1 General observations

The mathematical expression of the generalized coherent states depends on the creation / annihilation operators we chose, respectively on the result of their action on the Fock basis vectors. Therefore, there can be multiple variants of coherent states corresponding to the same quantum system. In the above we have chosen the pair of operators $\hat{\mathcal{A}}_+$ and $\hat{\mathcal{A}}_-$, whose ordered normal product is just equal to the dimensionless Hamiltonian of the system, $\hat{\mathcal{H}} = \# \hat{\mathcal{A}}_+ \hat{\mathcal{A}}_- \#$. In this case the mathematical expression of the energy eigenvalues, $e(n)$ , plays a decisive role. But, if instead of the operators $\hat{\mathcal{A}}_+$ and $\hat{\mathcal{A}}_-$ we choose another set of operators, for example, the generators of the quantum group associated with the system studied ($\hat{\mathcal{K}}_\pm$, $\hat{\mathcal{K}}_3$ for SU(1,1), or $\hat{S}_\pm$, $\hat{S}_3$ for SU(2) of spin systems), then other expressions for coherent states will be obtained. However, all these expressions will satisfy Klauder's conditions imposed on coherent states, so that they can also be included in the family, generically called generalized hypergeometric coherent states.

Next, we will present, in summary, some main results regarding generalized hypergeometric coherent states, constructed using the pair of operators indicated above, $\hat{\mathcal{A}}_+$ and $\hat{\mathcal{A}}_-$, for some cases of quantum oscillators.

To achieve this, the following steps must be taken:

- First of all, it is necessary to know the eigenvalues of the energy, as a result of solving the Schrodinger equation for stationary states will result in the expression of the energy eigenvalues $E_n = \hbar \omega_p e_q(n)$.

- The concrete expression of the dimensionless eigenvalues $e(n)$ will be compared with the most general expression. From here we will identify the coefficients $p$ and $q$, the sets of numbers $\{a_i\}_1^p$ , $\{A_i\}_1^p$ and $\{b_j\}_1^q$, $\{B_j\}_1^q$ so the expressions for the structure constant ${}_p\rho_q(n)$ and the normalization function ${}_pF_q(|z|^2)$ will result.

- Regardless of the definition, generalized coherent states will have the expression

$$|z>=\frac{1}{\sqrt{{}_pF_q\left(|z|^2\right)}}\sum_{n=0}^{n_{\max}\leq\infty}\frac{z^n}{\sqrt{{}_p\rho_q(n)}}|n>=$$

$$=\frac{1}{\sqrt{{}_pF_q\left(\left\{\left(\frac{a_i}{A_i}\right)_n\right\}_1^p;\left\{\left(\frac{b_j}{B_j}\right)_n\right\}_1^q;\frac{\prod_{i=1}^p A_i}{\prod_{j=1}^q B_j}|z|^2\right)}}\sum_{n=0}^{n_{\max}\leq\infty}\sqrt{\frac{\prod_{i=1}^p\left(\frac{a_i}{A_i}\right)_n}{\prod_{j=1}^q\left(\frac{b_j}{B_j}\right)_n}}\frac{\left(\sqrt{\frac{\prod_{i=1}^p A_i}{\prod_{j=1}^q B_j}}z\right)^n}{\sqrt{n!}}|n> \tag{6.1}$$

- Depending on the type of coherent state, in the above relation the order of the indices and constants will be: $p$ and $q$ , $\left\{\frac{a_i}{A_i}\right\}_1^p$ and $\left\{\frac{b_j}{B_j}\right\}_1^q$ for Barut-Girardello coherent states, respectively $q$ and $p$ , $\left\{\frac{b_j}{B_j}\right\}_1^q$ and $\left\{\frac{a_i}{A_i}\right\}_1^p$, for Klauder-Perelomov coherent states. Accordingly, the integration measure will be particularized from the common expression.

- The expected mean values will then be calculated, as well as the Mandel parameter.

- Finally, for mixed (thermal) states, the expressions for the distribution functions $P$ and $Q$ will be expressed, in the representation of generalized coherent states, as well as the thermal Mandel parameter.

In the calculations, the following relations will be useful

$$(x)_{n,y}=x(x+y)(x+2y)\ldots(x+(n-1)y)=\frac{\Gamma(x+y\,n)}{\Gamma(x)}=y^n\left(\frac{x}{y}\right)_n=y^n\frac{\Gamma\left(\frac{x}{y}+n\right)}{\Gamma\left(\frac{x}{y}\right)} \tag{6.2}$$

$$\lim_{y=0}y^n\left(\frac{x}{y}\right)_n=x^n$$

We quote for each case the characteristic parameters, the structure constants, the normalization function, and the expression of the generalized coherent states. For this we will write the expressions of the dimensionless energy eigenvalues ${}_pe_q(n)$, respectively of the structure constant ${}_p\rho_q(n)$, in the manner

$${}_pe_q(n)=n\frac{\prod_{j=1}^q\left[b_j+B_j(n-1)\right]}{\prod_{i=1}^p\left[a_i+A_i(n-1)\right]}\equiv n\;{}_pf_q(n)\quad;\quad{}_p\rho_q(n)=n!\frac{\prod_{j=1}^q(B_j)^n\left(\frac{b_j}{B_j}\right)_n}{\prod_{i=1}^p(A_i)^n\left(\frac{a_i}{A_i}\right)_n}\equiv n!\;{}_p\tilde{\rho}_q(n) \tag{6.3}$$

Using the above limit, we can formulate the following three types of limits, which are useful in deducing the expressions of generalized coherent states, corresponding to different types of energy spectra:

a) ***For systems whose energy dimensionless eigenvalues are expressible by a polynomial of order 1*** in the variable $n$, i.e. $\tilde{f}(n) = n + e_0$. These systems have only one pair of non-zero values in the set of numbers $\{a_i, A_i\}$ namely $a_1 = 1$, $A_1 = 1$, the remainder being zero. Consequently, the following limits applies:

$$\lim_{\substack{a_i=1 \\ A_i=1}} {}_1e_q(n) = \lim_{\substack{a_i=1 \\ A_i=1}} n \frac{\prod_{j=1}^{q}\left[b_j + B_j(n-1)\right]}{a_1 + A_1(n-1)} = \prod_{j=1}^{q}\left[b_j + B_j(n-1)\right] = {}_0e_q(n) \tag{6.4}$$

$$\lim_{\substack{a_i=1 \\ A_i=1}} {}_1\rho_q(n) = \prod_{s=1}^{n} {}_0e_q(n) = \prod_{j=1}^{q}\prod_{s=1}^{n}\left[b_j + B_j(s-1)\right] = \prod_{j=1}^{q}\left[b_j\right]\left[b_j + B_j\right]\ldots\left[b_j + B_j(n-1)\right] =$$
$$= \prod_{j=1}^{q}\frac{\Gamma(b_j + B_j n)}{\Gamma(b_j)} = \prod_{j=1}^{q}(B_j)^n\left(\frac{b_j}{B_j}\right)_n \tag{6.5}$$

$$\lim_{\substack{a_i=1 \\ A_i=0}} {}_1F_q\left(\left\{\frac{a_i}{A_i}\right\}_1^p ; \left\{\frac{b_j}{B_j}\right\}_1^q ; \frac{\prod_{i=1}^{p}(A_i)}{\prod_{j=1}^{q}(B_j)}|z|^2\right) =$$
$$= \sum_{n=0}^{\infty}\left[\lim_{\substack{a_i=1 \\ A_i=0}}\left(\frac{a_1}{A_1}\right)_n (A_1)^n\right]\frac{1}{\left(\frac{b_j}{B_j}\right)_n}\frac{\left(\frac{1}{\prod_{j=1}^{q}(B_j)}|z|^2\right)^n}{n!} = {}_0F_q\left( ; \left\{\frac{b_j}{B_j}\right\}_1^q ; \frac{1}{\prod_{j=1}^{q}(B_j)}|z|^2\right) \tag{6.6}$$

$${}_0F_1\left( ; \nu+1 ; \frac{x^2}{4}\right) = \Gamma(\nu+1)\left(\frac{x}{2}\right)^{-\nu} I_\nu(x) \tag{6.7}$$

$$(-x)_n = (-1)^n\frac{\Gamma(x+1)}{\Gamma(x+1-n)} \tag{6.8}$$

***Examples:***

***- Linear Harmonic Oscillator (HO-1D).*** *Characteristics:*

$$p = 0 \;;\; q = 0 \;;\; {}_0e_0(n) = n \;,\; {}_0\rho_0(n) = n! \;;\; {}_0F_0( \;;\; ; |z|^2) = e^{|z|^2} \;;\; |z\rangle = e^{-|z|^2}\sum_{n=0}^{\infty}\frac{z^n}{\sqrt{n!}}|n\rangle$$

***- The set of systems with energy spectrum a polynomia of the first degree*** $_0e_1(n)=b_1+B_1(n-1)$

$_0e_1(n)=C+n$

*Characteristics:*

$$p=0,\ a_1=1\ ,\ A_1=1\ ;\ q=1\ ,\ b_1=C+1\ ,\ B_1=1$$

$$_0e_1(n)=n+C\ ,\ _0\tilde{\rho}_1(n)=(C+1)_n$$

$$_0F_1(\ ;\ C+1\ ;|z|^2\ )=\sum_{n=0}^{\infty}\frac{1}{(C+1)_n}\frac{(|z|^2)^n}{n!}=\Gamma(C+1)|z|^{-C}I_C(2|z|)$$

$$|z>=\sqrt{\frac{|z|^C}{\Gamma(C+1)I_C(2|z|)}}\sum_{n=0}^{\infty}\frac{1}{\sqrt{(C+1)_n}}\frac{z^n}{\sqrt{n!}}|n>$$

| ***Quantum system*** | ***Characteristics*** | | | | |
|---|---|---|---|---|---|
| | $p$ $\{a_1\ ,\ A_1\}$ | $q$ $\{b_1\ ,\ B_1\}$ | $C$ | $_pf_q(n)$ | $_p\tilde{\rho}_q(n)$ |
| **Pseudoharmonic oscillator (PHO)** [Popov, 2001] | $p=0$ $a_1=1$ $A_1=1$ | $q=1$ $b_1=k+1$ $B_1=1$ | $k\ ,$ $k>0$ | $n+k$ | $(k+1)_n$ |
| **Meixner oscillator (Mx)** [Atakishiyev, 1998], [Popov, 2023], | $p=0$ $a_1=1$ $A_1=1$ | $q=1$ $b_1=\frac{\beta+2}{2}$ $B_1=1$ | $\frac{1}{2}\beta\ ,$ $\beta>0$ | $n+\frac{1}{2}\beta$ | $\left(\frac{\beta+2}{2}\right)_n$ |
| **An electron in a uniform magnetic field (e-)** [Dong, 2007] | $p=0$ $a_1=1$ $A_1=1$ | $q=1$ $b_1=\frac{m+1}{2}$ $B_1=1$ | $\frac{m+1}{2}\ ,$ $m>0$ | $n+\frac{m+1}{2}$ | $\left(\frac{m+3}{2}\right)_n$ |
| **Non interacting spin S systems in a uniform magnetic field (Spin)** [Radcliffe, 1971] [Popov, 2016] | $p=0$ $a_1=1$ $A_1=1$ | $q=1$ $b_1=1-S$ $B_1=1$ | $-S$ | $-S+n$ | $(1-S)_n$ |

Finally, the corresponding coherent states are:

**(PHO)**

$$|z>=\sqrt{\frac{|z|^k}{\Gamma(k+1)I_k(2|z|)}}\sum_{n=0}^{\infty}\frac{1}{\sqrt{(k+1)_n}}\frac{z^n}{\sqrt{n!}}|n>$$

**(Mx)** $$|z\rangle = \sqrt{\frac{|z|^{\frac{1}{2}\beta}}{\Gamma\left(\frac{1}{2}\beta+1\right) I_{\frac{1}{2}\beta}(2|z|)}} \sum_{n=0}^{\infty} \frac{1}{\sqrt{\left(\frac{1}{2}\beta+1\right)_n}} \frac{z^n}{\sqrt{n!}} |n\rangle$$

**(e⁻)** $$|z\rangle = \sqrt{\frac{|z|^{\frac{m+1}{2}}}{\Gamma\left(\frac{m+1}{2}+1\right) I_{\frac{m+1}{2}}(2|z|)}} \sum_{n=0}^{\infty} \frac{1}{\sqrt{\left(\frac{m+1}{2}+1\right)_n}} \frac{z^n}{\sqrt{n!}} |n\rangle$$

**(Spin)** $$|z\rangle = \sqrt{\frac{|z|^{-S}}{\Gamma(1-S) I_{-S}(2|z|)}} \sum_{n=0}^{\infty} \frac{1}{\sqrt{(1-S)_n}} \frac{z^n}{\sqrt{n!}} |n\rangle$$

***- The set of systems with energy spectrum a polynomial with degree two of the form*** $_0\tilde{f}(n)_2 = [b_1 + B_1(n-1)]\,[b_2 + B_2(n-1)] = B_1 B_2 \left(\frac{b_1}{B_1} - 1 + n\right)\left(\frac{b_2}{B_2} - 1 + n\right)$.

*Characteristics:* $p = 0,\; a_1 = 1\;,\; A_1 = 1\;;\; q = 2$

$$_0\tilde{f}_2(n) = \prod_{j=1}^{2} \left[b_j + B_j(n-1)\right] = (b_1 - B_1 + B_1 n)(b_2 - B_2 + B_2 n)$$

$$_0\tilde{\rho}_2(n) = \prod_{s=1}^{n} {}_0e_2(s) = \prod_{j=1}^{2}\prod_{s=1}^{n} \left[b_j + B_j(s-1)\right] = \prod_{j=1}^{2} (B_j)^n \left(\frac{b_j}{B_j}\right)_n =$$

$$= (B_1 B_2)^n \left(\frac{b_1}{B_1}\right)_n \left(\frac{b_2}{B_2}\right)_n$$

$$_0F_2\left(\;;\; \frac{b_1}{B_1}, \frac{b_2}{B_2}\;;\; \frac{1}{B_1 B_2}|z|^2\right) = \sum_{n=0}^{\infty} \frac{1}{\left(\frac{b_1}{B_1}\right)_n \left(\frac{b_2}{B_2}\right)_n} \frac{\left(\frac{1}{B_1 B_2}|z|^2\right)^n}{n!}$$

$$|z\rangle = \frac{1}{\sqrt{_0F_2\left(\;;\; \frac{b_1}{B_1}, \frac{b_2}{B_2}\;;\; \frac{1}{B_1 B_2}|z|^2\right)}} \sum_{n=0}^{\infty} \frac{1}{\sqrt{\left(\frac{b_1}{B_1}\right)_n \left(\frac{b_2}{B_2}\right)_n}} \frac{\left(\frac{1}{\sqrt{B_1 B_2}}\, z\right)^n}{\sqrt{n!}} |n\rangle$$

| Quantum system | Characteristics | | | | | |
|---|---|---|---|---|---|---|
| | $p$ $\{a_1\ ,\ A_1\}$ | $q$ $\{b_1\ ,\ B_1\}$ | $\frac{b_1}{B_1}$ | $\frac{b_2}{B_2}$ | ${}_p\tilde{f}_q(n)$ | ${}_p\tilde{\rho}_q(n)$ |
| **Morse oscillator (MO)** [Dong, 2007] | $p=0$ $a_1=1$ $A_1=0$ | $q=2$ $b_1=\frac{3-\nu}{2}$ $b_2=\frac{3-\nu}{2}$ $B_1=B_2=1$ | $\frac{3-\nu}{2}$ | $\frac{3-\nu}{2}$ | $\left(n-\frac{\nu-1}{2}\right)^2$ | $\left[\left(\frac{3-\nu}{2}\right)_n\right]^2$ |
| **Pöschl-Teller oscillator (PT)** [Pöschl, 1933] | $p=0$ $a_1=1$ $A_1=0$ | $q=2$ $b_1=1-s$ $b_2=1-s$ $B_1=B_2=1$ | $-s$ | 1 | $(n-s)^2$ | $[(1-s)_n]^2$ |
| **Pöschl-Teller hyperbolic (PT II)** [Martinez, 2024], | $p=0$ $a_1=1$ $A_1=0$ | $q=2$ $b_1=b_2=\frac{3-\lambda}{2}$ $B_1=B_2=1$ | $-\frac{\lambda-1}{2}$ | 1 | $\left(n-\frac{\lambda-1}{2}\right)^2$ | $\left[\left(\frac{3-\lambda}{2}\right)_n\right]^2$ |
| **Kerr medium (Kerr)** [Rohith, 2015] [] | $p=0$ $a_1=1$ $A_1=0$ | $q=2$ $b_1=1$ $B_1=1$ | 1 | $\frac{\omega}{\chi}$ | $n\left[\left(1-\frac{\chi}{\omega}\right)+\frac{\chi}{\omega}n\right]$ | $\left(\frac{\chi}{\omega}\right)^n(1)_n\left(\frac{\chi}{\omega}\right)_n$ |
| **Pöschl-Teller trigonometric (PT- trig)** [Antoin, 2001] | $p=0$ $a_1=1$ $A_1=0$ | $q=2$ $b_1=1$ $b_2=\nu+1$ $B_1=B_2=1$ | 1 | $\nu+1$ | $n(n+\nu)$ | $(1)_n(\nu+1)_n$ |
| **Infinitely deep square well-potential (IQW)** [Dong, 2007] [Popov, 2014] | $p=0$ $a_1=1$ $A_1=0$ | $q=2$ $b_1=b_2=1$ $B_1=B_2=1$ | 1 | 1 | $n^2$ | $(n!)^2$ |

Generally, the coherent states expression's is

$$|\,z> = \frac{1}{\sqrt{{}_0F_2\left(\;;\;\frac{b_1}{B_1},\frac{b_2}{B_2}\;;\;\frac{1}{B_1B_2}|\,z\,|^2\right)}}\sum_{n=0}^{\infty}\frac{1}{\sqrt{\left(\frac{b_1}{B_1}\right)_n\left(\frac{b_2}{B_2}\right)_n}}\frac{\left(\frac{1}{\sqrt{B_1B_2}}z\right)^n}{\sqrt{n!}}|\,n>$$

**(MO)** $$|\,z> = \frac{1}{\sqrt{{}_0F_2\left(\;;\;\frac{3-\nu}{2},\frac{3-\nu}{2}\;;\;|\,z\,|^2\right)}}\sum_{n=0}^{\infty}\frac{1}{\left(\frac{3-\nu}{2}\right)_n}\frac{z^n}{\sqrt{n!}}|\,n>$$

**(PT)** $$|\,z> = \frac{1}{\sqrt{{}_0F_2\left(\;;1-s\,,1-s\;;\;\frac{1}{c_2}|\,z\,|^2\right)}}\sum_{n=0}^{\infty}\frac{1}{(1-s)_n}\frac{\left(\frac{1}{\sqrt{c_2}}z\right)^n}{\sqrt{n!}}|\,n>$$

**(PT II)** $$|\,z> = \frac{1}{\sqrt{{}_0F_2\left(\;;\;\frac{3-\nu}{2},\frac{3-\nu}{2}\;;\;|\,z\,|^2\right)}}\sum_{n=0}^{\infty}\frac{1}{\left(\frac{3-\nu}{2}\right)_n}\frac{z^n}{\sqrt{n!}}|\,n>$$

**(Kerr)** $$|\,z> = \frac{1}{\sqrt{{}_0F_2\left(\;;\;1\,,\frac{\omega}{\chi}\;;\;\frac{\omega}{\chi}|\,z\,|^2\right)}}\sum_{n=0}^{\infty}\frac{1}{\sqrt{(1)_n\left(\frac{\omega}{\chi}\right)_n}}\frac{\left(\sqrt{\frac{\omega}{\chi}}\,z\right)^n}{\sqrt{n!}}|\,n>$$

**(PT- trig)** $$|\,z> = \frac{1}{\sqrt{{}_0F_2\left(\;;\;1\,,\nu+1\;;\;|\,z\,|^2\right)}}\sum_{n=0}^{\infty}\frac{1}{\sqrt{(1)_n(\nu+1)_n}}\frac{z^n}{\sqrt{n!}}|\,n>$$

**(IQW)** $$|\,z> = \frac{1}{\sqrt{{}_0F_2\left(\;;1,1\;;\;|\,z\,|^2\right)}}\sum_{n=0}^{\infty}\frac{1}{(1)_n}\frac{z^n}{\sqrt{n!}}|\,n>$$

c) ***For systems whose energy dimensionless eigenvalues are expressible by the inverse of a polynomial*** in the variable $n$, up to order 2.,

$$_0\tilde{f}(n)_2 = \frac{1}{[a_2+A_2(n-1)]\,[b_3+B_3(n-1)]} = \frac{1}{A_2A_3\left(\frac{a_2}{A_2}-1+n\right)\left(\frac{a_3}{A_3}-1+n\right)}\,.$$

*Characteristics:* $p=0$, $a_1=1$ , $A_1=1$ ; $q=2$ , $b_j=1$ , $B_j=0$ , $j=1,2,...,q$

$$ {}_0\tilde{\rho}(n)_2 = \prod_{s=1}^{n} {}_0\tilde{f}(n)_2 = \frac{1}{(A_2A_3)^n}\prod_{s=1}^{n}\frac{1}{\left(\frac{a_2}{A_2}-1+s\right)\left(\frac{a_3}{A_3}-1+s\right)} = \frac{1}{(A_2A_3)^n}\frac{1}{\left(\frac{a_2}{A_2}\right)_n\left(\frac{a_3}{A_3}\right)_n} $$

$$ {}_2F_0\left(\left(\frac{a_2}{A_2}\right)_n , \left(\frac{a_3}{A_3}\right)_n ; \quad ; A_2A_3|z|^2\right) = \sum_{n=0}^{\infty}\left(\frac{a_2}{A_2}\right)_n\left(\frac{a_3}{A_3}\right)_n\frac{\left(A_2A_3|z|^2\right)^n}{n!} $$

$$ |z> = \frac{1}{\sqrt{{}_2F_0\left(\left(\frac{a_2}{A_2}\right)_n , \left(\frac{a_3}{A_3}\right)_n ; \quad ; A_2A_3|z|^2\right)}}\sum_{n=0}^{\infty}\sqrt{\left(\frac{a_2}{A_2}\right)_n\left(\frac{a_3}{A_3}\right)_n}\frac{\left(\sqrt{A_2A_3}\,z\right)^n}{\sqrt{n!}}|n> $$

$$ \lim_{\substack{a_i=1\\ A_i=1\\ b_{j\neq 2}=1\\ B_{j\neq 2}=0}} {}_pF_q\left(\left\{\frac{a_i}{A_i}\right\}_1^p ; \left\{\frac{b_j}{B_j}\right\}_1^q ; \frac{\prod_{i=1}^{p}(A_i)}{\prod_{j=1}^{q}(B_j)}|z|^2\right) = {}_{p-1}F_0\left(\left\{\frac{a_i}{A_i}\right\}_2^p ; \quad ; \prod_{i=2}^{p}(A_i)|z|^2\right) \qquad () $$

| ***Quantum system*** | ***Characteristics*** | | | | | |
|---|---|---|---|---|---|---|
| | $p$<br>$\{a_1\ ,\ A_1\}$ | $q$<br>$\{b_1\ ,\ B_1\}$ | $\frac{a_2}{A_2}$ | $\frac{a_3}{A_3}$ | ${}_2\tilde{f}_0(n)$ | ${}_2\tilde{\rho}_0(n)$ |
| **Kratzer oscillator (Krat)** [Kratzer, 1920] | $p=2$<br>$a_1=1$<br>$A_1=1$<br>$a_2=\varepsilon$<br>$A_2=1$<br>$a_3=\varepsilon$<br>$A_3=1$ | $q=0$<br>$b_j=1$ ,<br>$B_j=0$ ,<br>$j=1,2,\ldots,q$ | $\varepsilon$ | $\varepsilon$ | $\frac{1}{(n+\varepsilon)^2}$ | $\frac{1}{[(\varepsilon+1)_n]^2}$ |
| **Hydrogen-like atoms (H)** [Messiah, 1967] | $p=2$<br>$a_1=1$<br>$A_1=1$<br>$a_2=1$<br>$A_2=1$<br>$a_3=1$<br>$A_3=1$ | $q=0$<br>$b_j=1$ ,<br>$B_j=0$ ,<br>$j=1,2,\ldots,q$ | 1 | 1 | $\frac{1}{n^2}$ | $\frac{1}{(n!)^2}$ |

$$| z >= \frac{1}{\sqrt{{}_2F_0\left(\left(\frac{a_2}{A_2}\right)_n , \left(\frac{a_3}{A_3}\right)_n ; \; ; A_2A_3 | z |^2\right)}} \sum_{n=0}^{\infty} \sqrt{\left(\frac{a_2}{A_2}\right)_n \left(\frac{a_3}{A_3}\right)_n} \frac{\left(\sqrt{A_2A_3}\, z\right)^n}{\sqrt{n!}} | n >$$

**(Krat)**

$$| z >= \frac{1}{\sqrt{{}_2F_0\left(\varepsilon , \varepsilon ; \; ; | z |^2\right)}} \sum_{n=0}^{\infty} \sqrt{(\varepsilon)_n (\varepsilon)_n} \frac{z^n}{\sqrt{n!}} | n >$$

**(H)**

$$| z >= \frac{1}{\sqrt{{}_2F_0\left(1 , 1; \; ; | z |^2\right)}} \sum_{n=0}^{\infty} \sqrt{(1)_n (1)_n} \frac{z^n}{\sqrt{n!}} | n >$$

## 7. Concluding remarks

After the concept of coherent states appeared (although not under this name) exactly 100 years ago, for three or four decades it did not arouse much interest from the scientific community. Only in the second half of the last century, a series of renowned physicists (let's just mention the most notable ones, Glauber, Klauder and Perelomov) drew attention to the fact that coherent states have real application potential, especially in the field of quantum optics. As expected, perhaps also under the influence of Schrodinger's original paper, coherent states were initially formulated for the one-dimensional harmonic oscillator. The justification was simple: this ideal model allows the testing of all new theoretical concepts, especially due to the mathematical facilities it involves and which lead to conclusive and clear results / conclusions.

Later, the question was raised whether coherent states could also be defined for anharmonic oscillators, which, in many cases, more adequately describe the behavior of physical systems than the harmonic model. As a concession, however, certain initial requirements had to be deliberately given up, namely that coherent states minimize uncertainty relations. In this way, the family of coherent states has been continuously enriched with new "members", including anharmonic oscillators of various types (pseudoharmonic, Morse, Poschl-Teller and so on), as well as other quantum systems and phenomena (hydrogenoid atoms, Kerr effect, Landau levels).

In this paper we did not aim to present an exhaustive history of the development of coherent states over a century (i.e. a paper like Dodonov's [Dodonov, 2002]). This type of approach would be relatively difficult, given the almost exponential development of the field of coherent states (along with closely related fields, such as squeezed states, photon added and photon contracted states and so on). Therefore, we preferred to start by examining the most general coherent states - generalized hypergeometrical coherent states (direction introduced by Appl and Schiller [Appl, 2004]), because generalized hypergeometrical functions are among the most general special functions from which an extremely large number of common special functions can be deduced, as particular cases (which, in turn, can serve as normalized functions for the different types of coherent states). Inspired by the normally ordered operator integration technique, IWOP, introduced by Fan [Fan, 2003] , applicable to the harmonic oscillator, we have developed a generalization of it, applicable to anharmonic oscillators - Diagonal Operators Ordering Technique (DOOT) [Popov, 2015b]. This has proven very useful due to the results regarding anharmonic oscillators. In conclusion, let us emphasize that, although in most applications canonical coherent states (corresponding to the linear harmonic oscillator, the linear dependence being only a first approximation for many phenomena) are used, we express our belief that in the not-so-distant future, as more and more nonlinear characteristics will be taken into account for the examination of quantum phenomena, generalized coherent states will be increasingly involved. In passing, let's list just some of the fields in which coherent states have already found their purpose: Quantum information, Molecular dynamics, Cosmology, Field theory and so on.